\documentclass[aps, prd, twocolumn, groupedaddress, nofootinbib]{revtex4-2}

\usepackage{amsmath}
\usepackage{amssymb}
\usepackage{graphicx}
\usepackage{mathtools} 
\usepackage{lipsum}
\usepackage{xcolor}
\usepackage[hidelinks, colorlinks=true, citecolor=blue, linkcolor=black, urlcolor=blue]{hyperref}
\usepackage{cleveref}
\usepackage[export]{adjustbox}
\usepackage{subcaption}
\usepackage{booktabs}               

\usepackage{tikz} 

\usetikzlibrary{arrows.meta} 
\usetikzlibrary{backgrounds} 
\usetikzlibrary{positioning} 
\usetikzlibrary{fit} 
\usetikzlibrary{calc} 
\usetikzlibrary{patterns} 
\usetikzlibrary{fadings} 
\usetikzlibrary{decorations.pathmorphing} 
\usetikzlibrary{decorations.pathreplacing} 
\usetikzlibrary{decorations.markings} 
\usetikzlibrary{shapes} 
\usetikzlibrary{shapes.misc} 
\usetikzlibrary{math}
\usetikzlibrary{arrows}
\tikzset{cross/.style={cross out, draw, 
minimum size=2*(#1-\pgflinewidth), 
         inner sep=0pt, outer sep=0pt}}

\tikzset{snakeit/.style={decorate, decoration=snake}}

\newcommand{\drawvertexrad}[3]{
    \draw[very thick, fill=white] (#1,#2) circle[radius=#3];
    \draw[very thick, pattern=north west lines] (#1,#2) circle[radius=#3];
}

\newcommand{\drawvertexempty}[3]{
    \draw[very thick, fill=white] (#1,#2) circle[radius=#3];
    \draw[very thick, fill=black] (#1,#2) circle[radius=#3*0.2];
}

\newcommand{\selfenergyquarticcoupling}[1]{
    \tikzmath{\unit=#1;}
    \begin{tikzpicture}[scale=\unit, baseline={([yshift=-.5ex]current bounding box.center)}]
        \draw[very thick] (-2,0) -- (0,0);
        \draw[very thick] (0,0) -- (2,0);
        \draw[very thick] (0,1) circle[radius=1];
        \draw[very thick, fill=black] (0, 0) circle[radius=0.125];
    \end{tikzpicture}
}

\newcommand{\selfenergysignoperatorannotations}[1]{
    \tikzmath{\unit=#1;}
    \begin{tikzpicture}[scale=\unit, baseline={([yshift=-.5ex]current bounding box.center)}]
        \draw[very thick] (-2,0) -- (0,0);
        \draw[very thick] (0,0) -- (2,0);
        \draw[very thick] (0,1) circle[radius=1];
        \drawvertexempty{0}{0}{#1};
        \node (upper) at (0, 2.4) {$q$};
        \node (left) at (1.5,0.4) {$p$};
        \node (right) at (-1.5,0.4) {$p$};
    \end{tikzpicture}
}

\newcommand{\onetothreescatter}[1]{
    \tikzmath{\unit=#1;}
    \begin{tikzpicture}[scale=\unit, baseline={([yshift=-.5ex]current bounding box.center)}]
        \draw[very thick] (-2, 0) -- ( 0, 0);
        \draw[very thick] ( 0, 0) -- ( 2, 0);
        \draw[very thick] ( 0, 0) -- ( 2, 1);
        \draw[very thick] ( 0, 0) -- ( 2,-1);
        \draw[very thick, fill=black] (0, 0) circle[radius=0.125];
        \node (A) at (-2.4, 0) {$\phi$};
        \node (B) at ( 2.4, 0) {$\phi$};
        \node (C) at ( 2.4, 1.2) {$\phi$};
        \node (D) at ( 2.4,-1.2) {$\phi$};
    \end{tikzpicture}
}

\newcommand{\bubble}[1]{
    \tikzmath{\unit=#1;}
    \begin{tikzpicture}[scale=\unit, baseline={([yshift=-.5ex]current bounding box.center)}]
        \draw[very thick] (0,0) ellipse (1.5 and 1);
        \draw[very thick] (-2,-1.0) -- (-1.5,0);
        \draw[very thick] (-2, 1.0) -- (-1.5,0);
        \draw[very thick] ( 2,-1.0) -- ( 1.5,0);
        \draw[very thick] ( 2, 1.0) -- ( 1.5,0);
        \draw[very thick, fill=black] (-1.5,0) circle[radius=0.125];
        \draw[very thick, fill=black] (1.5,0) circle[radius=0.125];
    \end{tikzpicture}
}

\newcommand{\bubbleannotations}[1]{
    \tikzmath{\unit=#1;}
    \begin{tikzpicture}[scale=\unit, baseline={([yshift=-.5ex]current bounding box.center)}]
        \draw[very thick] (0,0) ellipse (1.5 and 1);
        \draw[very thick] (-2,-1.0) -- (-1.5,0);
        \draw[very thick] (-2, 1.0) -- (-1.5,0);
        \draw[very thick] ( 2,-1.0) -- ( 1.5,0);
        \draw[very thick] ( 2, 1.0) -- ( 1.5,0);
        \draw[very thick, fill=black] (-1.5,0) circle[radius=0.125];
        \draw[very thick, fill=black] (1.5,0) circle[radius=0.125];
        \node (upper) at (0, 1.4) {$q$};
        \node (lower) at (0,-1.4) {$q+p$};
        \node (leftup)  at (-2.1,1.3) {$p_1$};
        \node (leftdown)  at (-2.1,-1.3) {$p_2$};
        \node (rightup) at (2.1,1.3) {$p_3$};
        \node (rightdown) at (2.1,-1.3) {$p_4$};
    \end{tikzpicture}
}

\newcommand{\fourvertexNOLABEL}[1]{
    \tikzmath{\unit=#1;}
    \begin{tikzpicture}[scale=\unit, baseline={([yshift=-.5ex]current bounding box.center)}]
        \draw[very thick] (-2.0,-2.0) -- ( 2.0, 2.0);
        \draw[very thick] (-2.0, 2.0) -- ( 2.0,-2.0);
        \drawvertexrad{0}{0}{1}
    \end{tikzpicture}
}
\newcommand{\fourvertexpointlikeNOLABEL}[1]{
    \tikzmath{\unit=#1;}
    \begin{tikzpicture}[scale=\unit, baseline={([yshift=-.5ex]current bounding box.center)}]
        \draw[very thick] (-2.0,-2.0) -- ( 2.0, 2.0);
        \draw[very thick] (-2.0, 2.0) -- ( 2.0,-2.0);
        \draw[very thick, fill=black] (0,0) circle[radius=#1];
    \end{tikzpicture}
}

\newcommand{\fourvertexsignoperatorNOLABEL}[1]{
    \tikzmath{\unit=#1;}
    \begin{tikzpicture}[scale=\unit, baseline={([yshift=-.5ex]current bounding box.center)}]
        \draw[very thick] (-2.0,-2.0) -- ( 2.0, 2.0);
        \draw[very thick] (-2.0, 2.0) -- ( 2.0,-2.0);
        \drawvertexempty{0}{0}{1}
    \end{tikzpicture}
}

\newcommand{\genericvertex}[1]{
    \tikzmath{\unit=#1;}
    \begin{tikzpicture}[scale=\unit, baseline={([yshift=-.5ex]current bounding box.center)}]
        \draw[very thick] (-2.0,-2.0) -- ( 0.0, 0.0);
        \draw[very thick] (-2.0, 2.0) -- ( 0.0, 0.0);
        \draw[very thick, snakeit] ( 0.0, 0.0) -- ( 2.0, 2.0) node[right]{$\mu_1\nu_1$};
        \draw[very thick, snakeit] ( 0.0, 0.0) -- ( 2.0,-2.0) node[right]{$\mu_n\nu_n$};
        \def \pointshift{0.4};  
        \draw[very thick, fill=black] (-2.0+\pointshift, 0.75) circle[radius=0.0125];
        \draw[very thick, fill=black] (-2.0+\pointshift, 0.0) circle[radius=0.0125];
        \draw[very thick, fill=black] (-2.0+\pointshift,-0.75) circle[radius=0.0125];
        \draw[very thick, fill=black] ( 2.0-\pointshift, 0.75) circle[radius=0.0125];
        \draw[very thick, fill=black] ( 2.0-\pointshift, 0.0) circle[radius=0.0125];
        \draw[very thick, fill=black] ( 2.0-\pointshift,-0.75) circle[radius=0.0125];
        \drawvertexrad{0}{0}{0.5}
    \end{tikzpicture}
}

\newcommand{\iu}{\mathrm{i}} 
\newcommand{\eu}{\mathrm{e}}
  
\newcommand{\dd}{\mathrm{d}}

\begin{document}

\title{Shear viscosity of a relativistic scalar field from functional renormalization}

\author{Tim Stoetzel}
\email[]{tim.stoetzel@uni-jena.de}
\affiliation{Theoretisch Physikalisches Institut, Friedrich-Schiller-Universität Jena, Max-Wien-Platz 1, 07743 Jena, Germany}

\author{Stefan Floerchinger}
\email[]{stefan.floerchinger@uni-jena.de}
\affiliation{Theoretisch Physikalisches Institut, Friedrich-Schiller-Universität Jena, Max-Wien-Platz 1, 07743 Jena, Germany}

\begin{abstract}
Renormalization group flow equations of the fluid dynamical shear viscosity transport coefficient of a relativistic real scalar field are derived. 
The flowing effective action contains branch cut contributions to the self energy and interaction vertex in the symmetric phase. 
We demonstrate how the flow equation method can systematically extend the perturbative resummation schemes. 
We show that our truncation is in that sense a minimal scheme in which a reliable viscosity coefficient is obtained.

\end{abstract}


\maketitle


\section{Introduction}
Relativistic fluid dynamics has been successfully employed in a variety of fields, ranging from cosmology and astrophysics~\cite{Weinberg:2008zzc,Ghiglieri:2015nfa,Ma:1995ey,Abbott:1984fp,birrell_davies_1982,Calzetta:1986ey,Caprini:2018mtu,Drewes:2023oxg,Flauger:2017ged,Hawking:1966qi,Mazumdar_2019,rezzolla_relativistic_2013} down to microscopic systems such as heavy ion collisions~\cite{annurev.nucl.54.070103.181236,Calzetta:2008iqa,doi:10.1142/S0217751X13400113,Floerchinger:2021xhb,Schafer:2009dj,Romatschke:2017ejr}. 
A quark-gluon plasma produced in high-energy collisions, can be described effectively using relativistic fluid dynamics. 
Deviations from local thermal equilibrium lead to dissipative effects such as damping of kinetic motion through viscosities or heat conduction~\cite{Romatschke:2007mq,Heinz:2013th}.
Commonly, these dissipative effects are systematically incorporated by employing a gradient expansion for the energy momentum tensor and conserved currents, up to a specific order.
The expansion introduces new parameters which specify the efficiency of equilibration processes like heat or momentum transport~\cite{ISRAEL1979341,Mueller1967,landau1987fluid,Kovtun:2012rj,rezzolla_relativistic_2013}. 
While fluid dynamics predicts the existence of these transport coefficients, it does not give any information on their dependence on temperature or chemical potential. 
Information of these quantities must be found either by experiments or first principle calculations from microscopic descriptions, since small-scale interactions govern ultimately also the macroscopic transport in the fluid.
This leads to the interesting question how effective, dissipative theories can be constructed, starting from a microscopic description. 

From a field-theoretic point of view, transport coefficients can be accessed through Kubo relations~\cite{Kubo:1957mj,Kubo:1957wcy} if the fluid is still close to its equilibrium state. 
These relations involve real-time correlation functions of conserved currents. 
Several approaches have been developed to compute transport coefficients, using for example Lattice field theory~\cite{Meyer:2009jp}, kinetic theory approaches~\cite{Ghiglieri:2018dib,Arnold:2003zc,Enss:2012nx}, perturbative QFT~\cite{Jeon_1992,Jeon_1995} or effective theories in the description of cold quantum gases~\cite{Enss:2019ydh,Enss:2010qh}.
While these methods can evaluate correlators in specific regimes, such as the high-temperature or high-density regimes, it is particularly interesting to consider how transport properties can be determined within non-perturbative methods, as for example considered in refs.~\cite{Christiansen:2014ypa,Lowdon:2021ehf,Rose:2016elj,Rose:2017lui}. 
The goal is to address interacting quantum field theories, such as strongly interacting quantum chromodynamics (QCD), in all relevant regimes.

This work is concerned with the calculation of transport coefficients, using non-perturbative functional renormalization group methods~\cite{Wilson:1973jj,DUPUIS2021,Reuter_Saueressig_2019,Gies:2006wv, Berges:2000ew}. 
In order to show how dissipative dynamics can be described, and which microscopic processes need to be accounted for, we will take a relativistic $\varphi^4$ theory of massive particles as an example.
We show, as a proof of principle, how a shear viscosity coefficient in this rather simple theory can be calculated as a function of the temperature.
The latter is introduced with the Matsubara or imaginary time formalism~\cite{Matsubara:1955ws}.

The shear viscosity coefficient is extracted from the real-time correlation function using the analytic continuation formalism for effective actions, developed in refs~\cite{Floerchinger:2011sc,Floerchinger:2016gtl} and also used in the context of scalar field theories in refs~\cite{Jung:2021ipc,Kamikado:2013sia,Rose:2017lui,Rose:2016elj}.
The methods allows to introduce dissipative transport properties as imaginary terms in the flowing action, i.e. branch cuts in the frequency dependence. 
We emphasize, that these terms are not only extracted from the flow but feed back into it, and this is crucial for the determination of the shear viscosity.
For example, the scalar propagator develops a thermal broadening during the RG flow at finite temperature, which itself has an important influence on other flow equations.
This scheme allows to overcome a problematic feature of perturbation theory at finite temperature~\cite{Bros:2001zs,Landsman:1988ta,Lowdon:2024atn,Lowdon:2024pcb}.

This paper is structured as follows:
In \cref{ch:phi4_and_viscosity} the relativistic $\varphi^4$ model is introduced at finite temperature together with the description of dissipative fluids.
A connection is made between the macroscopic and microscopic theories using linear response theory by the use of a Kubo formula.
Section \ref{ch:FRG} introduces the FRG methods and the flow equation. 
We furthermore show how dissipative terms in the effective action are constructed and give further information about the propagator and interaction vertices of the effective field theory.
Afterwards the calculation of the evolution equations of the shear viscosity and other observables is shown in \cref{ch:flowequations}, especially focusing on the evaluation of the Matsubara sums and the analytic continuation to real time observables.
Numerical results for the viscosity coefficient and the Landau damping are shown in \cref{ch:results}.
Technical details are provided in several appendices.

\textit{Conventions \& Notation}: We are working in units where $k_\mathrm{B}=\hbar=c=1$ and furthermore use the Minkowski spacetime with signature ($-,+,+,+$). 
We furthermore introduce the abbreviations 
\begin{equation}
    \int_x = \int \dd^4 x, \, \int_{\mathbf{p}} = \int \frac{\dd^3 \mathbf{p}}{(2\pi)^3},\, \sum_{\omega_n} = T\sum_{n=-\infty}^\infty.
\end{equation}

\section{Real Scalar Field QFT \& Shear Viscosity}
\label{ch:phi4_and_viscosity}
\subsection{Microscopic model}
We consider a real and relativistic scalar field with $\mathbb{Z}_2$ symmetry. 
Because we are interested in correlation functions of the energy-momentum tensor, it is convenient to use general coordinates and consider the real time microscopic action by a functional of the field $\varphi(x)$ as well as the (externaly provided) spacetime metric $g_{\mu\nu}(x)$,
\begin{equation}
    \begin{split}
        S[\varphi; g] = \int \dd^4 x \sqrt{g(x)} \bigg\{& -\frac{1}{2 } g^{\mu\nu}(x)\partial_\mu \varphi (x)\partial_\nu \varphi(x) \\
        &- \frac{m^2}{2}\varphi(x)^2 - \frac{\lambda }{4! }\varphi(x)^4 \bigg\}.
    \end{split}
    \label{eq:phi4_microscopic}
\end{equation}
We are interested in the symmetric phase and assume a positive squared mass $m^2>0$ and positive quartic coupling $\lambda$. We use the abbreviation $g(x) = -\det (g_{\mu\nu}(x))$.  

Let us recall here that a variation of the action with respect to the spacetime metric yields the symmetric energy-momentum tensor, $\delta S = \int \dd^4 x \sqrt{g(x)} (1/2) \hat T^{\mu\nu}(x) \delta g_{\mu\nu}(x)$. 
The quantum expectation value of this operator plays an important role for the macroscopic description of the theory in terms of fluid dynamics.

We use the standard methods of the functional integral formulation of quantum field theory, such as the generating functional of connected correlation functions or Schwinger functional $W[J;g]$, defined through
\begin{equation} 
    \eu^{\iu W[J;g]} = \int \mathcal{D}\varphi \, \eu^{\iu S[\varphi;g] + \iu\int_x J(x)\varphi(x)},
    \label{eq:schwinger_functional_definition}
\end{equation}
with the abbreviation $\smash{\int_x = \int \dd^4 x \sqrt{g(x)}}$. 
Connected correlation functions are obtained by taking derivatives of $W[J;g]$ with respect to the external source $J(x)$ and evaluating these at the point $J=0$. 

\begin{figure}[htbp]
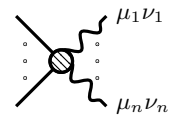

    \centering   
    \genericvertex{0.3}
    \caption{The $\Gamma^{(m,n)\mu_1\nu_1\ldots\mu_n\nu_n}$ vertex's diagrammatic representation.
            Each solid line represents an external scalar field, while a wavy line corresponds to an external metric field.}
    \label{fig:generic_vertex_diagram}
\end{figure}

We furthermore introduce the one-particle irreducible (1PI) quantum effective action $\Gamma[\phi, g]$ which is the generating functional of correlation function that include all quantum corrections and is defined as the Legendre transform of $W[J]$,
\begin{equation} 
    \Gamma[\phi; g] = \sup_J \int_x J(x) \phi(x) - W[J[\phi], g],
\end{equation}
where $\phi(x) = \langle \varphi(x)\rangle$ is the scalar field expectation value. 
Equation of motion for the field expectation values can be obtained from the quantum effective action with
\begin{equation}
    \frac{\delta \Gamma[\phi; g] }{\delta \phi(x) } = J(x).
\end{equation}
Higher-order functional derivatives yield effective vertices or 1PI correlation functions.
For the remainder of this work we use the shorthand notation 
\begin{equation}
    \begin{split}
        &\Gamma^{(m, n)\mu_{1}\ldots \mu_{2n}}[\phi; g](x_1,\ldots x_m; y_1,\ldots y_n) = \\
        &\frac{\delta^{m+n} \Gamma[\phi; g]}{\delta \phi(x_1)\cdots \delta \phi(x_m)\delta g_{\mu_1\mu_2}(y_1)\ldots g_{\mu_{2n-1}\mu_{2n}}(y_n)},
    \end{split} 
\end{equation}   
and any $\Gamma^{(m,n)}$ vertex will be diagrammatically represented as in \cref{fig:generic_vertex_diagram}.

A finite temperature $T>0$ can be introduced via the imaginary-time or Matsubara formalism~\cite{Matsubara:1955ws}. 
We will briefly recall the concepts, for a more detailed introduction see refs \cite{LeBellac_1996, Kapusta_Gale_2006, Laine:2016hma, Abrikosov:107441}. 
Even though we started with a formulation for general coordinates in \cref{eq:phi4_microscopic}, our calculations will eventually be done in Minkowski spacetime. 
To calculate the grand-canonical partition function, we perform a Wick rotation of the time coordinate to $t= -\iu \tau $ with imaginary time $\tau$. 
The latter is compacted onto the interval $[0,\beta]$, and we impose periodic boundary conditions such that $\varphi(t-\iu\beta, \mathbf{x}) = \varphi(t,\mathbf{x})$. 
This accounts for thermal and quantum fluctuations in a bosonic theory. 
After this procedure, all fields now depend on the imaginary time $\tau$ and in momentum space this leads to the discrete Matsubara frequencies $\omega_n = 2\pi n T$ where $n\in \mathbb{Z}$.

Real-time observables can be extracted from the imaginary time formalism with an analytic continuation back onto the real time or frequency axis. 
The analytic continuation of correlation functions will be discussed in more detail in \cref{ch:flowequations}.

We also add an ultraviolet cutoff $\Lambda$ for the spatial momentum to circumvent the well known triviality problem of scalar $\varphi^4$ theory in four spacetime dimensions. 
Within perturbation theory at finite order, the problem can be solved with counterterms, but a non-perturbative renormalization scheme does not allow to fully remove the ultraviolet regularization of the momentum integrals at finite interaction strength, see for example refs.~\cite{Berges:2000ew,Gies:2006wv}.

\subsection{Relativistic fluid dynamics}
Relativistic fluid dynamics is an effective theory that describes interacting quantum fields when either the fluid velocity, or the typical velocity of constituents are comparable with the speed of light.
It is based on a close-to-local equilibrium approximation that is applicable to systems with many particles or quantum field excitations with a local equilibrium time $\tau_\text{eq}$ that is short compared to the typical dynamical time scales. 

Assuming there are no conserved currents besides energy and momentum, a relativistic fluid can be described by the energy momentum tensor $T^{\mu\nu}(x)$. 
It obeys the covariant conservation law 
\begin{equation}
    \nabla_\mu T^{\mu\nu}(x) = 0,
    \label{eq:tmunu_conservationlaw}
\end{equation}
which can be seen to be a consequence of invariance under general coordinate transformations or diffeomorphisms.
We work in the Landau frame where the energy-momentum current gets decomposed as 
\begin{equation}
    T^{\mu\nu} = \epsilon u^\mu u^\nu + (p+\Pi ) \Delta^{\mu\nu} + \pi^{\mu\nu}.   
    \label{eq:tmunu_definition}
\end{equation}
Here, $u^\mu$ is the four-velocity of a fluid element, defined to be the time-like eigenvector of the energy-momentum tensor, with energy density $\epsilon$ being the corresponding eigenvalue, and $p$ is the thermodynamic pressure, that is related to $\epsilon$ via a thermodynamic equation of state that can be calculated from a microscopic description in thermal equilibrium through standard methods.

The fluid velocity has normalization $u_\mu u^\mu = -1$ and $\Delta^{\mu\nu} = g^{\mu\nu} + u^\mu u^\nu$ defines an orthogonal projector with $\Delta^{\mu\nu}u_\nu = 0$.

For an ideal fluid one would have vanishing bulk viscous pressure $\Pi$ and shear stress $\pi^{\mu\nu}$, but beyond the ideal fluid approximations these are governed by dissipative effects.
The bulk pressure and shear stress capture processes that transport momentum internally while transferring kinetic energy from the macroscopic fluid motion to internal energy. 
Microscopically this happens e.g. due to particle collision processes out of detailed balance.

In order to close the fluid equations of motion, additional relations must be provided for the bulk viscous pressure and shear stress. 
This is usually done in terms of a derivative expansion. 
To first order in gradients, or within 'first order hydrodynamics', these constitutive equations are
\begin{equation}
\begin{split}
    \Pi =& -\zeta \nabla_\mu u^\mu, \\
    \pi^{\mu\nu} =& -2\eta \Delta^{\mu\nu\phantom{\alpha}\beta}_{\phantom{\mu\nu}\alpha}\nabla_\beta u^\alpha,
\end{split} \label{eq:constitutive_eq}
\end{equation}
where $\smash{\Delta^{\mu\nu\phantom{\alpha}\beta}_{\phantom{\mu\nu}\alpha} = \frac{1}{2}\left( \Delta^\mu_{\phantom{\mu}\alpha} \Delta^{\nu\beta} + \Delta^{\mu\beta}\Delta^\nu_{\phantom{\nu}\alpha}  \right) - \frac{1}{3}\Delta^{\mu\nu}\Delta_\alpha^{\phantom{\alpha}\beta}}$. The coefficients $\eta$ and $\zeta$ are the shear- and bulk viscosities.
The transport coefficients in general depend on temperature $T$, and other thermodynamic variables like chemical potentials, and must be determined from a close-to-equilibrium calculation in the microscopic theory. 
It matters here what constituents the fluid is made of and how they interact with each other on small time and length scales compared to the macroscopic time and length scales.

A bridge between the macroscopic fluid theory, and a microscopic description in terms of a quantum field theory can be made in terms of correlation and response functions, as we recall next. 

\subsection{Linear Response Theory}
Response theory, and especially linear response theory~\cite{Kubo:1957mj,Kubo:1957wcy} allows us to connect a microscopic theory, in our case the scalar quantum field theory in eq.\ \eqref{eq:phi4_microscopic}, to a macroscopic theory, which is in our case relativistic fluid dynamics described by eqs.\ \eqref{eq:tmunu_conservationlaw}, \eqref{eq:tmunu_definition} and \eqref{eq:constitutive_eq}.

More specifically, linear response theory can be used to answer the question of how the energy momentum tensor evolves in a given spacetime geometry in a region close to its equilibrium configuration.
Let us consider the decomposition of the spacetime metric $g_{\mu\nu}(x) = \eta_{\mu\nu} + \delta g_{\mu\nu}(x)$ with a small deviation $\delta g_{\mu\nu}(x)$ around the Minkowski spacetime. 
A fluctuation in the background naturally induces a fluctuation in the energy-momentum tensor with $T^{\mu\nu}(x) = \bar T^{\mu\nu} + \delta T^{\mu\nu}(x)$, where $\bar T^{\mu\nu}$ is a background configuration that we take to be a homogenous equilibrium state in Minkowski space, $\bar T^{\mu\nu} = (\bar\epsilon + \bar p) \bar u^{\mu} \bar u^{\nu} + \bar p \eta^{\mu\nu}$. 

The retarded response function is defined through
\begin{equation}
    G_\mathrm{R}^{\mu\nu\rho\sigma}(x,y) = \frac{4}{\sqrt{g(x)}\sqrt{g(y)}} \frac{\delta}{\delta g_{\rho\sigma}(y)} \frac{1}{2}\sqrt{g(x)} T^{\mu\nu}(x),
    \label{eq:def_retarded_response}
\end{equation}
so that for small deviation, $\left|\delta g_{\mu\nu}\right|\ll 1$,
\begin{equation}
    \delta \left(\sqrt{g(x)} T^{\mu\nu}(x)\right) = \frac{1}{2}\int\dd^4 y  G_\mathrm{R}^{\mu\nu\rho\sigma}(x,y) \delta g_{\rho\sigma}(y).
    \label{eq:def_linear_response_general}
\end{equation}
Note that as a consequence of relativistic causality, the retarded response function is vanishing except when $x$ is in the future of $y$.
As a consequence of translational symmetry of the global equilibrium state, the response function only depends on the coordinate differences and \cref{eq:def_linear_response_general} becomes 
\begin{equation}
    \delta \left(\sqrt{g(x)} T^{\mu\nu}(x)\right) = \frac{1}{2}\int\dd^4 y  G_\mathrm{R}^{\mu\nu\rho\sigma}(x-y) \delta g_{\rho\sigma}(y).
    \label{eq:def_linear_response}
\end{equation}

The response function $G_\mathrm{R}^{\mu\nu\alpha\beta}(x-y)$ encodes information about the instantaneous response connected to the thermodynamic properties of the fluid as well as information about relaxation processes that drive the fluid back into equilibrium. 
The latter can be used to find relations between different correlation functions of the microscopic fields and the transport coefficients.

We define the Fourier representation of the response function in Minkowski space as 
\begin{equation}
\begin{split}
    G_\mathrm{R}^{\mu\nu\alpha\beta}(x-y) =&\\
    \int \frac{\dd \omega \dd^3 \mathbf{k} }{(2\pi)^4 }& \eu^{-\iu\omega (x^0-y^0) + \iu\mathbf{k}(\mathbf{x}-\mathbf{y})}G_\mathrm{R}^{\mu\nu\alpha\beta}(\omega,\mathbf{p}),
\end{split}
\end{equation}
Linearizing the covariant conservation law \eqref{eq:tmunu_conservationlaw} and the constitutive relations \eqref{eq:constitutive_eq} around the Minkwoski background, one can find Kubo formulas which relate the transport coefficients to the response function.
Specifically for the shear viscosity, one relation is given by
\begin{equation}
    \eta = -\lim_{\omega\rightarrow 0}\frac{\partial}{\partial \omega} \mathrm{Im} \, G_\mathrm{R}^{xyxy}(\omega,\mathbf{0}).
    \label{eq:kubo_formula}
\end{equation}
For a general derivation of transport coefficient see \cite{Kadanaoff_Martin_1963,stoetzel2025energymomentumresponsemetricperturbations,Kovtun:2012rj}. 
Particularly for the shear viscosity Kubo formula we refer the reader to refs~\cite{Jeon_1992,Czajka:2017,Jeon:2025,Moore:2010bu}. 
We would also like to point out that this is only one possible way to extract the shear viscosity, as discussed in ref~\cite{Jeon:2025}.

Equation \eqref{eq:kubo_formula} shows that the shear viscosity coefficient is located in the imaginary part of the purely spatial, off-diagonal entries of the response kernel. 
The microscopic theory needs to account for this by supplying the correct analytic structure in complex frequency plane. 
Since the energy momentum tensor can be calculated from the effective action using 
\begin{equation}
    T^{\mu\nu}(x) = \frac{2 }{\sqrt{g}(x)} \frac{\delta \Gamma[\phi; g]}{\delta g_{\mu\nu}(x)},
\end{equation}
we can combine \cref{eq:def_retarded_response} and \cref{eq:kubo_formula} and calculate the viscosity from the retarded two-point function.
Note that the energy-momentum tensor correlation function is defined by 
\begin{equation}
    G^{\mu\nu\alpha\beta}(x-y) = 4\frac{\delta^2 \Gamma[\phi; g]}{\delta g_{\mu\nu}(x)\delta g_{\alpha\beta}(y)}.
    \label{eq:def_responsefunction_effective_action}
\end{equation}
Evaluating this expression on the Matsubara-torus yields the energy-momentum tensor correlation function at imaginary time $\tau$. 
This allows us to define the momentum space representation 
\begin{equation}
    G^{\mu\nu\alpha\beta}(x-y) = \sum_{\omega_n} \int_{\mathbf{k}} \eu^{-\iu\omega_n(x^0 - y^0) + \iu\mathbf{k}(\mathbf{x}-\mathbf{y})}G^{\mu\nu\alpha\beta}(\iu\omega_n, \mathbf{k})
\end{equation}
with $x^0,y^0$ being imaginary times and the Matsubara frequency $\omega_n = 2\pi n T$. 
The knowledge of the imaginary time correlator at the Matsubara frequencies is sufficient to determine all other correlation functions, like the retarded or advanced correlators~\cite{Altland_Simons_2010,LeBellac_1996,Kapusta_Gale_2006}.
The correlation function can be analytically continued from the Matsubara frequencies to the complex plane with $\iu\omega_n\rightarrow z$, except for the real axis where possible poles and branch cuts might appear. 
Any other correlation function can be constructed from the continued correlator $G^{\mu\nu\alpha\beta}(z,\mathbf{p})$, by approaching the real frequency axis from a suitable direction.
For example, the retarded correlation function, or response function, can be obtained by analytically continuing the frequency onto the real line from above. 
Using $\iu\omega_n \rightarrow \omega + \iu\epsilon$ with $\omega$ being real and $\epsilon > 0$ one finds the response function 
\begin{equation}
    G^{\mathrm{R}\,\mu\nu\alpha\beta}(\omega,\mathbf{k}) = \lim_{\epsilon\rightarrow 0} G^{\mu\nu\alpha\beta}(\omega + \iu \epsilon, \mathbf{k}).
    \label{eq:tmunu_correlator_analytic_continuation}
\end{equation} 

\section{Functional Renormalization Group Methods}
\label{ch:FRG}
The Kubo formula \eqref{eq:kubo_formula} suggests that the shear viscosity $\eta$ can be calculated from the effective action $\Gamma[\phi, g]$, using the analytic continuation given in \cref{eq:tmunu_correlator_analytic_continuation}.
Perturbative methods would achieve this by summing up all relevant Feynman diagrams up to a certain order in the coupling or some other small parameter and renormalizing the theory introducing counter terms. 
We choose the non-perturbative FRG scheme.
Instead of expanding around a small coupling the method builds upon the idea of the Wilsonian renormalization group~\cite{Wilson:1971bg,Wilson:1973jj}, by consecutively integrating out fluctuations of different energy scales, starting from the microscopic scales up to an infrared scale. 
The FRG methods build up on an evolution equation that governs this coarse graining and results in the quantum effective action. 
We will only give a short introduction here, while further information can be found in the established literature, see refs~\cite{Reuter_Saueressig_2019, DUPUIS2021, Gies:2006wv, Berges:2000ew, Kopietz:2010zz}.

\subsection{Effective Average Action}
We start by adding an infrared regulator term to the microscopic action $S[\varphi, g]$ as 
\begin{equation}
    S_k[\varphi; g] = S[\varphi; g] + \Delta S_k[\varphi; g],
\end{equation}
where $\smash{\Delta S_k[\varphi; g] = \frac{1}{2}\int_{x,y} \varphi(x) R_k(x,y)\varphi(y)}$. 
The new parameter $k$ is the \textit{RG-scale}, which we take to be on the interval $[0, \Lambda]$.
The quantity $\Lambda$ is a UV scale which characterizes the microscopic dynamics trough which $S[\phi, g]$ is defined, and arises naturally if one does not take the continuum limit of the theory. 
In this work, we identify this UV cutoff with the cutoff of spatial-momentum integrals.
The cutoff could be taken to infinity for a theory that is renormalizable.
In our case we cannot do this since it is well known that $\varphi^4$ theory at non-vanishing coupling is non-perturbatively non-renormalizable in four spacetime dimensions and needs a UV regularization at non-vanishing interaction $\lambda$. 
The role of the regulator $R_k(x,y)$ here is to suppress specific fluctuations with momenta $p^2$ below the RG scale $k^2$. 
It is in general a function of $k$ that needs to vanish in the IR limit and single out fluctuations of $p^2=k^2$ in the ultraviolet. 
In our work we will restrict ourself to the Callan-Symanzik (CS) regulator defined by $\smash{R_k(x,y) = \delta^{(4)}(x-y) k^2}$. 
The advantage here is the locality and causality it imposes on the propagators since $k^2$ is effectively just an additive contribution to the mass squared and as such does not substantially alter the frequency and momentum dependence of the propagators.

The change of the microscopic theory in presence of the regulator gives rise to a scale dependent partition function $Z_k[J]$ and Schwinger functional $W_k[J, g]$ defined analogously to \cref{eq:schwinger_functional_definition}.
Furthermore one can define the effective average action  
\begin{equation}
    \Gamma_k[\phi; g] = \tilde{\Gamma}_k[\phi; g] - \Delta S_k[\phi; g],
\end{equation}
where $\tilde{\Gamma}_k$ is the Legendre transform of $W_k$.  
The effective average action coincides formally with $S[\phi;g]$ at $k=\Lambda$, up to one-loop terms suppressed at large $\Lambda$, while it equals $\Gamma[\phi;g]$ at $k=0$. 
It can thus be regarded as a functional that interpolates between the microscopic theory at the UV scale $\Lambda$ and the quantum effective action $\Gamma[\phi; g]$. 

\subsection{FRG Flow Equation}
While the effective average action defines the theory on some scale $k$ between the UV and the IR, one can study how a change in the scales influences the coarse grained theory given by $\Gamma_k[\phi; g]$. 
This evolution is described by the \textit{Wetterich equation} \cite{Wetterich_1993}
\begin{equation}
    \partial_t \Gamma_k[\phi; g] = \frac{1}{2}\text{STr}\left[ \frac{\partial_t R_k }{\Gamma_k^{(2,0)}[\phi; g] + R_k } \right],
    \label{eq:def_wetterich_equation}
\end{equation}
where we work with the \textit{RG-time} $t = \log\left( \Lambda/k \right)$ and STr refers to tracing over all algebraic open indices and integrating over all spacetime region or momenta in Fourier space.
Additionally, the equation is supplemented with the initial condition $\Gamma_{k=\Lambda}[\phi;g]=S[\phi;g]$.

The propagator of the scalar field $\phi$ at a scale $k$ is given by the denominator of \cref{eq:def_wetterich_equation} with $\smash{G_k^{-1} = \Gamma_k^{(2,0)}[\phi; g] + R_k}$.
Higher order correlators can be found by applying functional derivatives to both sides which results in flow equations of specific $n$-point functions. 

The flow equation \eqref{eq:def_wetterich_equation} is exact and describes the flow of the effective average action $\Gamma_k[\phi; g]$, coinciding with the microscopic theory $S[\phi; g]$ at $k=\Lambda$ and with the effective action $\Gamma[\phi; g]$ at $k=0$.
The evolution of $\Gamma_k[\phi; g]$ is in general regulator dependent, however $\Gamma_0[\phi;g]$ does not depend on the choice of the regulator $R_k$.
Furthermore, $\Gamma_k[\phi; g]$ depends on the full knowledge of $\smash{\Gamma_k^{(2,0)}}$ and similarly, the flow equation of any $n$-point function $\smash{\Gamma_k^{(n,0)}}$ depends on $\smash{\Gamma_k^{(n+1,0)}}$ and $\smash{\Gamma_k^{(n+2,0)}}$.
In order to manage this infinite tower of equations one usually truncates the space of solutions by formulating an ansatz for the effective average action. 
This produces a finite set of equations that can be solved, while also introducing an error in the IR theory which, in general, depends on the truncation of the effective average action and the regulator.
In this work, we will focus on a minimal truncation which is able to calculate a shear viscosity coefficient, motivated by the perturbative setup of ref~\cite{Jeon_1995}. 
More details on the ansatz are shown the next section. 

We can make use of the Wetterich equation \eqref{eq:def_wetterich_equation} and the Kubo formula defined by \cref{eq:kubo_formula} together with the analytic continuation \eqref{eq:tmunu_correlator_analytic_continuation} to formulate a flow equation of the shear viscosity coefficient.
Note that the energy-momentum tensor correlation function is defined analogously to equation \eqref{eq:def_responsefunction_effective_action} but is now scale dependent.
\begin{equation}
    \begin{split}
        \partial_t \eta_k = -\lim_{\omega\rightarrow 0}\frac{\partial }{\partial \omega }\mathrm{Im}\, \partial_t G_k^{\mathrm{R}\,xyxy}(\omega,\mathbf{0}),
    \end{split}
    \label{eq:eta_k_def}
\end{equation}
with the response function being defined via the analytic continuation of the imaginary time energy-momentum tensor correlation function $G_k^{\mu\nu\alpha\beta}(\iu\omega_n, \mathbf{k})$ analogously to the previous section. 
The temperature dependence of the shear viscosity can then be extracted by solving the flow equation \cref{eq:eta_k_def} for different temperatures up to $k=0$, once suitable initial conditions and truncations have been formulated.

\subsection{Truncation of the Effective Average Action}
While there are several ways to truncate the effective action systematically, we focus on an extension of the approach that is known as the derivative expansion. 
For a more general overview on truncation schemes see refs~\cite{DUPUIS2021,Gies:2006wv,Berges:2000ew,Kopietz:2010zz,Pawlowski:2005xe}.

The effective average action will be truncated at a certain order in the fields but also only including terms involving spacetime derivatives up to a specific order.
We restrict ourselves to an ansatz including fields up to the fourth order and derivatives up to order $\partial^2$. 
Since the effective average actions needs to respect the underlying $\mathbb{Z}_2$ symmetry, we only consider terms even in the fields.
This will leave us with a quartic interaction term and a kinetic term.
Additionally, we will also couple the action minimally to the external metric field which is used as a source for the response function. 

Furthermore, effectively dissipative or imaginary terms play an important role for us. 
Formally, \cref{eq:kubo_formula} suggests that we need terms in the effective action which couple to the external source quadratically, but also introduce a non-vanishing imaginary part in the complex frequency plane.
The imaginary part is related to branch cuts of the two-point function $\smash{\Gamma_k^{(0,2)xyxy}(\omega,\mathbf{p})}$ and usually corresponds to decay or production processes from interactions with the heat bath.

More generally, we will split the effective average action into a regular and a discontinuous part with  
\begin{equation}
    \Gamma_k[\phi; g] = \Gamma_k^\text{r}[\phi; g] + \Gamma_k^\text{d}[\phi; g], 
\end{equation}
where branch-cut-terms and the scale dependent viscosity coefficient $\eta_k$ are contained in the second term.

Starting with the regular part, we expand the formulation of the microscopic action by adding a scale dependent mass $m_k^2$ and keeping the coupling fixed.
We define
\begin{equation}
    \begin{split} 
        \Gamma_k^\mathrm{r}[\phi; g] = \int_x &\sqrt{g(x)}\bigg\{-\frac{1}{2}g^{\mu\nu}(x)\partial_\mu\phi(x) \partial_\nu \phi(x) \\
        &- \frac{m_k^2 }{2 }\phi^2 - \frac{\lambda }{4! }\phi^4 + \text{$g_{\mu\nu}\,$-terms}\bigg\}.
    \end{split}
    \label{eq:def_gamma_reg}
\end{equation}
The scale dependent mass is chosen to incorporate thermal corrections to the mass. 
Because our focus is a proof-of-concept study of medium effects, we neglect the contributions to the flow of $m_k$ that describe vacuum fluctuation effects.
The interaction strength $\lambda$ is fixed to avoid dealing with the triviality problem of interacting $\phi^4$ theories in four spacetime dimensions~\cite{Peskin:1995ev,WeinbergQM2:1996} and to keep our results comparable to the perturbative calculations in ref~\cite{Jeon_1995}. 
In general one can also add a wave function renormalization $Z_k(\phi)$ to the kinetic term. 
Here we have chosen $Z_k=1$ and neglect the corresponding flow. 
However, if one would add this coefficient to the action, this would effectively supply a resummation of the $\smash{\Gamma_k^{(2,1)}}$ vertex which is closely related to the bubble resummation of the shear viscosity diagram encountered in ref~\cite{Jeon_1995}.
The last term of \cref{eq:def_gamma_reg} contains contributions thermodynamic quantities like the thermodynamic pressure or susceptibilities which appear in the ideal fluid description but are not relevant for the shear viscosity coefficient. 
For a discussion of the general structure of these terms see ref~\cite{stoetzel2025energymomentumresponsemetricperturbations}.

The discontinuous terms all exhibit branch cuts in the complex frequency plane and are constructed using the sign-operator $\smash{\mathrm{s}_\mathrm{I}(\omega)}$, introduced in~\cite{Floerchinger_2016}.
This symbol can be defined in momentum space with 
\begin{equation}
    \mathrm{s}_\mathrm{I}(\omega) = \mathrm{sgn}(\mathrm{Im}\,\omega),
\end{equation}
where $\omega\in\mathbb{C}$, and it is used to implement the proper analytic structure of $n$-point functions.
The sign-operator becomes positive on the upper complex half-plane, while being negative in the lower half-plane.
Its real space representation is formally given by 
\begin{equation}
    \mathrm{s}_\mathrm{R}(\partial_t) = \mathrm{sgn}(\mathrm{Re}\,\partial_t).
\end{equation} 

With this, we define
\begin{equation}
    \begin{split}
        \Gamma_k^\text{d}[\phi; g] = \int_x \sqrt{g(x)}\bigg\{& \frac{\gamma_k }{2} \phi(x)s_\mathrm{R}(\partial_t)\partial_t \phi(x) \\
        &+ \frac{1}{8}\phi^2(x)\kappa_k(-\vec{\nabla}^2)s_\mathrm{R}(\partial_t)\partial_t\phi^2(x) \\
        &+ \text{$g_{\mu\nu}\,$-terms}\bigg\}.
    \end{split}
    \label{eq:def_gamma_dis}
\end{equation}
The first term in \cref{eq:def_gamma_dis} introduces Landau damping~\cite{landau_1965} by adding a friction coefficient to the two-point function.
While Landau damping usually describes damping of charged particles by an electric field, here the effect appears due to interactions of the scalar field with the heat bath environment. 
Particles can decay into the bath or be created from the environment and the effect is most dominant when $k\approx T$.
In momentum space, the sign operator $\mathrm{s}_\mathrm{I}(\omega)$ ensures that the retarded two-point function obeys causality by shifting the poles to the lower complex plane.
The second term introduces similarly a damping for the four-point function. 
Including this term in the truncation effectively introduces a two-loop structure in the one-loop flow equation \eqref{eq:def_wetterich_equation}.
While this term is linear in the frequency, the full dependence of the coupling on the spatial momentum is kept. 
The last term of \cref{eq:def_gamma_dis} contains contributions involving the transport coefficients like the viscosities.
We will give these terms in their explicit form in future works.

Note that in general $\Gamma_k^\mathrm{d}$ can be made covariant since we can use the fluids rest frame with velocity $u^\mu=(1,0)$ to define $\partial_t = u^\mu \partial_\mu = \mathcal{L}_u$ with the Lie-derivative $\mathcal{L}_u$ in the direction of the fluid velocity $u^\mu$. 
A similar truncation, also introducing a damping coefficient and an effective two-loop structure, was considered in the ref~\cite{Roth:2021nrd} using the closed-time-path formalism.

\subsection{Propagator}
Given the truncation defined in \cref{eq:def_gamma_reg,eq:def_gamma_dis}, we can have a look at the analytic structure of the propagator and vertex functions.
Starting with the former, we access the two-point correlation function with 
\begin{equation}
    \Gamma_k^{\mathrm{E}\,(2,0)}(x_1, x_2) = \frac{\delta^2 \Gamma_k[\phi; g ]}{\delta \phi(x_1)\delta\phi(x_2)}\bigg|_{\phi=0},
\end{equation}
evaluated on the Matsubara torus with $x_1^0,x_2^0\in[0,\beta]$.
The Fourier representation is introduced by 
\begin{equation}
    \begin{split}
        \Gamma_k^{\mathrm{E}\,(2,0)}&(x_1 - x_2) =\\
        &\sum_{\omega_n}\int_{\mathbf{p}} \eu^{-\iu\omega_n(x_1^0 - x_2^0) + \iu\mathbf{p}(\mathbf{x} - \mathbf{y})} P^\mathrm{E}_k(\iu\omega_n, \mathbf{p}),
    \end{split}
\end{equation}
with $\mathrm{E}$ referring to the Euclidean, imaginary time, correlation function.
Similar to the previous sections, the two-point correlation function evaluated at the Matsubara frequencies can be continued onto the complex plane, such that it is analytic and has no branch cuts and zero crossings except on the real line. 
This defines the complex-argument two-point function $P_k(z,\mathbf{p}) = P_k^\mathrm{E}(z,\mathbf{p})$ with $z\in\mathbb{C}$.
With this definition, we can define the analytically continued, scale-dependent, propagator 
\begin{equation}
    G_k^{-1}(z,\mathbf{p}) = P_k(z,\mathbf{p}) + k^2.
\end{equation}
The retarded propagator is obtained from this with $z \rightarrow \omega + \iu \epsilon$ with $\epsilon > 0$.
\begin{equation}
    G_k^{\mathrm{R}}(\omega,\mathbf{p}) = \lim_{\epsilon \rightarrow 0 } G_k(\omega + \iu\epsilon, \mathbf{p}).
\end{equation}
Within our truncation the propagator continued to the complex plane can be read off from \cref{eq:def_gamma_reg,eq:def_gamma_dis} 
\begin{equation}
    G_k^{-1}(z,\mathbf{p}) = -z^2 + \mathbf{p}^2 + m_k^2 + k^2 - \iu\gamma_k z\, \mathrm{s}_\mathrm{I}(z),
    \label{eq:inverse_propagator_truncation}
\end{equation}
where $\mathrm{s}_\mathrm{I}(z) = \mathrm{sgn}\left( \mathrm{Im}\, z \right)$.
\begin{figure}[htbp]
    \begin{subfigure}{0.5\columnwidth}
        \includegraphics[width=1.0\textwidth, trim=3.3cm 0cm 3.3cm 0cm, clip]{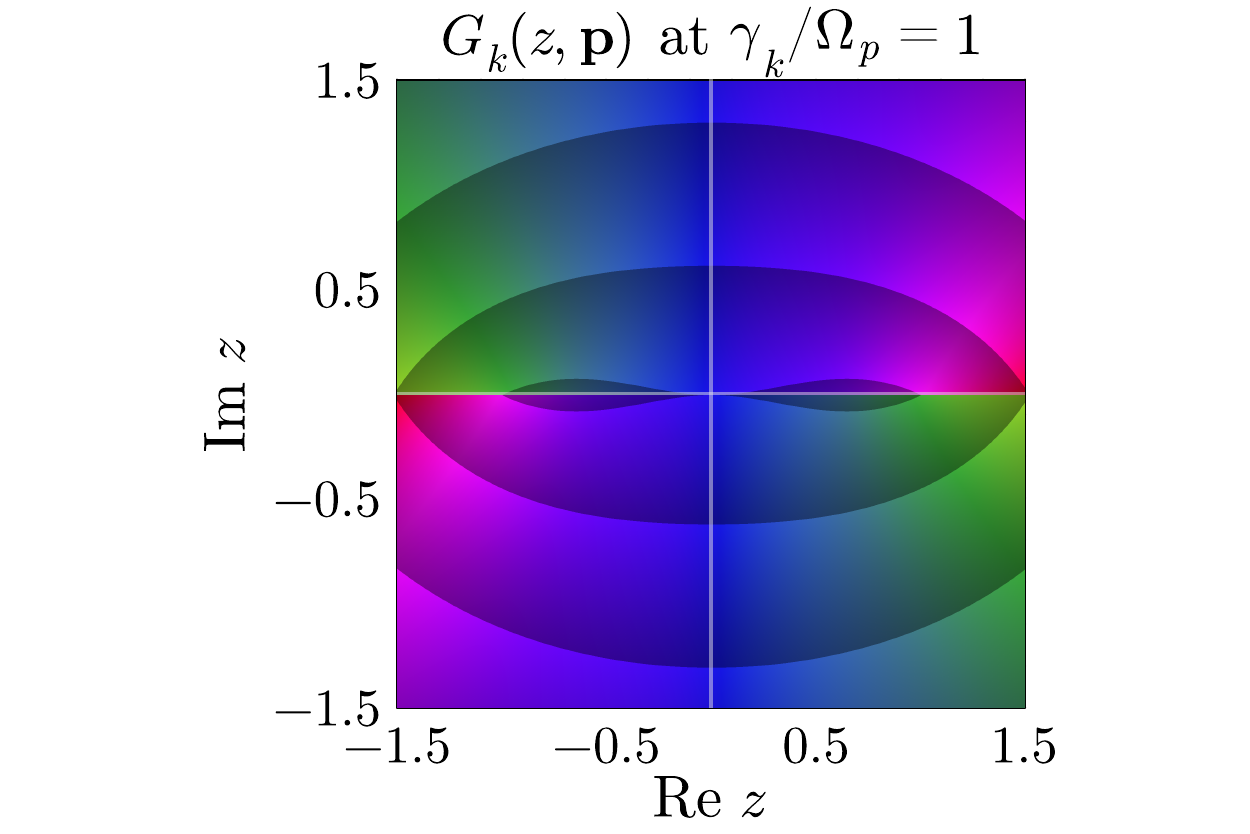}
    \end{subfigure}%
    \begin{subfigure}{0.5\columnwidth}
        \includegraphics[width=1.0\textwidth, trim=3.3cm 0cm 3.3cm 0cm, clip]{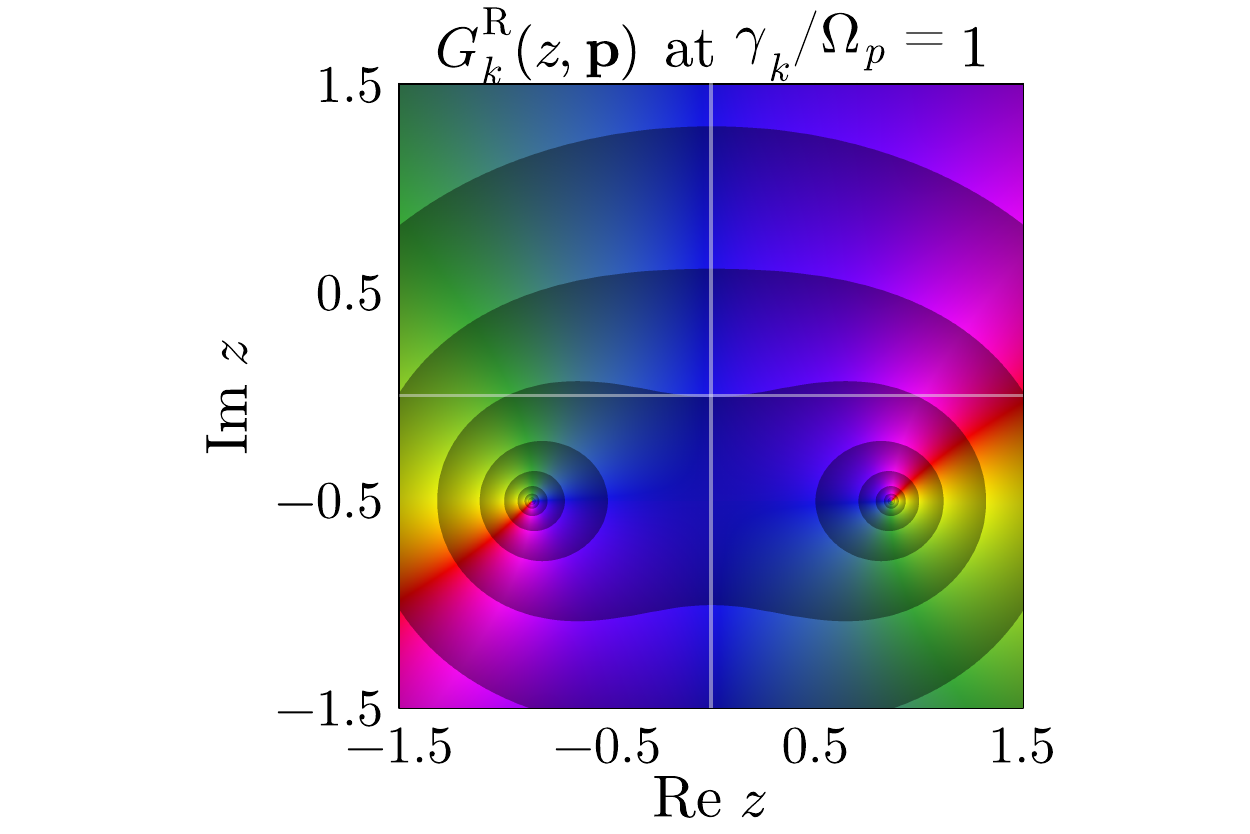}
    \end{subfigure}
    \caption{Analytic structure of the propagator given by \cref{eq:inverse_propagator_truncation} on the left and the retarded propagator on the right at finite $\Omega_p^2=m_k^2+k^2+\mathbf{p}^2$ and $\gamma_k$ for the complex frequency $z$.}
    \label{fig:propagator_poles}
\end{figure}
Without a choice of contour, the sign operator will introduce a branch cut on the real line, vanishing at $\mathrm{Re}\,z = 0$. 
The pole structure is fixed once the frequency is continued to the real line. 
In general, we see that a larger value of the Landau damping $\gamma_k$ shifts the poles further way onto the different Riemann-sheet and additionally enhances the branch cut generated by the sign-operator.
We do also see that critical damping would be achieved when $\smash{\gamma_k^2=4(m_k^2 + \mathbf{p}^2 + k^2)}$. 
At this specific value, both poles in the lower/upper half-plane merge into one point at the imaginary axis.

\subsection{Spectral density}
Using \cref{eq:inverse_propagator_truncation} we can also calculate the spectral function $\rho(\omega,\mathbf{k})$ by taking the retarded contour and extracting the imaginary part such that we find 
\begin{equation}
    \rho_k(\omega,\mathbf{p}) = 2\mathrm{Im}G_k^\mathrm{R}(\omega,\mathbf{p})= \frac{2\gamma_k \omega }{(\omega^2-\Omega_p^2)^2 + \gamma_k^2 \omega^2},
    \label{eq:spectral_function_truncation}
\end{equation}
where we introduced the abbreviation 
\begin{equation} 
    \Omega_p^2 = m_k^2 + \mathbf{p}^2 + k^2.
    \label{eq:dispersion_scale_dependent}
\end{equation}
If the damping coefficient $\gamma_k$ is smaller than $2\Omega_p$, the poles of the spectral density in the lower complex plane are located at 
\begin{equation}
    \begin{split}
        z_1(\mathbf{p}) =& -\sqrt{\Omega_p^2 - \frac{\gamma_k^2}{2} + \frac{\iu\gamma_k}{2}\sqrt{4\Omega_p^2-\gamma_k^2}},\\
        z_2(\mathbf{p}) =& \quad\,\sqrt{\Omega_p^2 - \frac{\gamma_k^2}{2} - \frac{\iu\gamma_k}{2}\sqrt{4\Omega_p^2-\gamma_k^2}}.        
    \end{split}
    \label{eq:spectral_density_poles_lower_plane}
\end{equation}
In the case where $\gamma_k > 2\Omega_p$, the poles in the lower complex plane are located at 
\begin{equation}
    \begin{split}
        z_1(\mathbf{p}) =& -\sqrt{\Omega_p^2 - \frac{\gamma_k^2}{2} + \frac{\gamma_k}{2}\sqrt{\gamma_k^2-4\Omega_p^2}}, \\
        z_2(\mathbf{p}) =& -\sqrt{\Omega_p^2 - \frac{\gamma_k^2}{2} - \frac{\gamma_k}{2}\sqrt{\gamma_k^2-4\Omega_p^2}},
    \end{split}
\end{equation}
and if $\smash{\gamma_k=2\Omega_p}$, only one double degenerate pole exists at $z=-\iu\Omega_p$ in the lower complex half plane. 
In all three cases, the two additional poles in the upper half plane are given by the conjugates $\smash{z_3=z_1^*}$ and $\smash{z_4=z_2^*}$.

\begin{figure}
    \includegraphics[width=1.0\columnwidth]{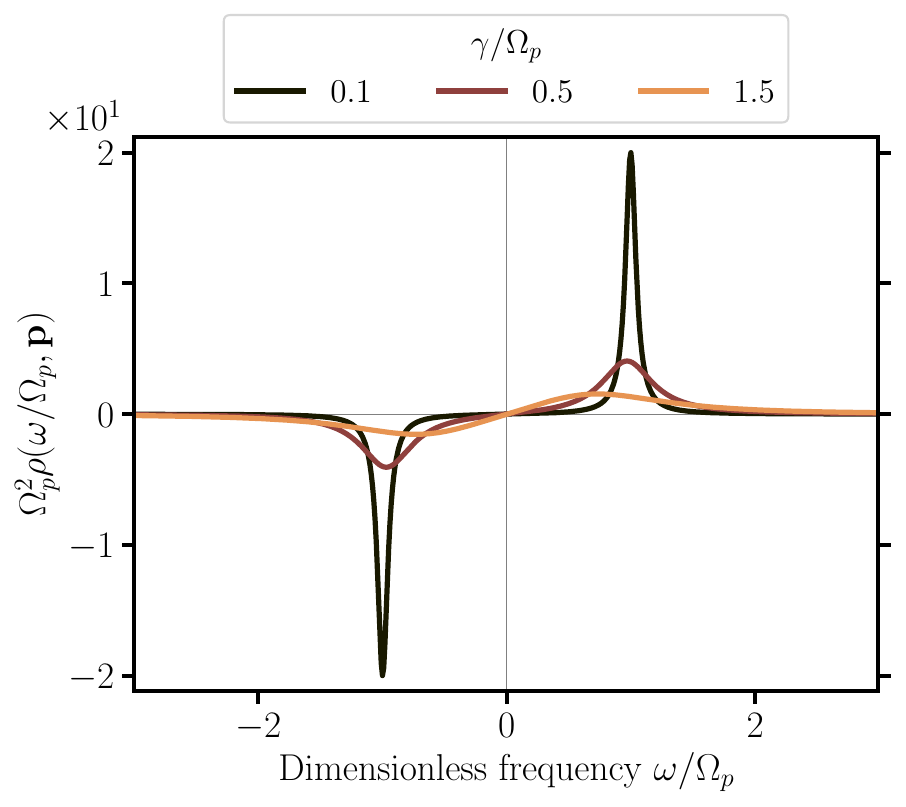}
    \caption{Shape of the spectral density for different ratios of $\gamma_k/\Omega_p$.}
    \label{fig:spectral_density_shape}
\end{figure}

The case $\gamma_k \ge 2\Omega_p$ does only arise in the regime of a strongly interacting theory at high temperature and since the Landau damping is usually small for all cases that are considered in this work, it suffices to work with \cref{eq:spectral_density_poles_lower_plane}.
Furthermore, if $\smash{\gamma_k \ll \Omega_p}$ we can use the 'pinching-pole' approximation, where the poles in the lower half-plane are characterized by the dispersion relation 
\begin{equation}
    z_{1,2}(\mathbf{p}) \approx \pm\Omega_p -\iu \frac{\gamma_k }{2}.
\end{equation}
In this context, $\gamma_k$ should be interpreted as a decay width of scalar particles.
The loss process appears due to interactions with the heat bath, allowing production or decay processes out-of or into the heat bath to which the system is coupled at finite temperature.
On a diagrammatic level this would allow for the decay process 
\begin{equation*}
    \onetothreescatter{0.5},
\end{equation*}
where the decaying scalar field can have an energy much lower than the production threshold, if at least one of the decay products is an excitation borrowed from the heat bath. 
This corresponds to a non-vanishing imaginary part of the response function which is necessary at low energies as suggested by the Kubo formula \eqref{eq:kubo_formula}.
Furthermore, one should note that \cref{eq:spectral_function_truncation} converges to a Dirac delta at low damping. 
Taking the limit $\gamma_k\rightarrow 0$ shows that we end up with the spectral density of the tree-level theory described by $S[\phi; g]$ with 
\begin{equation}
    \rho_k(\omega,\mathbf{p})\overset{\gamma_k\to 0 }{\longrightarrow }2\pi\mathrm{sgn}(\omega)\delta\left( \omega^2 - m_k^2 - k^2 - \mathbf{p}^2 \right),
    \label{eq:limit_of_spectral_density}
\end{equation}
which shows that the interactions with the thermal bath are just broadening the peak corresponding to the on-shell mass $m_k$. 
This behaviour is also depicted in \cref{fig:spectral_density_shape} for different damping coefficients. 
When the damping is small compared to the mass and momentum, the peak is located at the resonance frequency $\omega = \Omega_p$, while further increments of the damping coefficient shift the peak towards the origin.
At very large damping coefficients, where $\gamma_k~\gg~2\Omega_p$, the spectral density would appear to be only determined by one overdamped mode.
Eq. \eqref{eq:inverse_propagator_truncation} does also allow us to define the projectors for the squared mass and Landau damping with
\begin{align}
    m_k^2 =& \mathrm{Re}\lim_{\epsilon\rightarrow 0}\,P_k(\omega+\iu\epsilon, \mathbf{p})|_{p=0}, \label{eq:def_projector_mk}\\
    \gamma_k =& \frac{1}{2m_k }\mathrm{Im}\lim_{\epsilon\rightarrow 0} P_k(\omega + \iu\epsilon, \mathbf{p})|_{\omega=m_k, \mathbf{p}=0} \label{eq:def_projector_gammak}.
\end{align}
 
\subsection{Vertices}
The analytic structure of the four-point function does also contain a branch cut, see \cref{eq:def_gamma_dis}.
The discontinuous part of the vertex is expanded up to linear order in the frequency while the momentum dependence is kept fully, parameterized by $\smash{\kappa_k(\mathbf{p}^2)}$.
This choice is made because the 'bubble diagram' corresponding to a $\phi^2$-$\phi^2$ interaction,  cannot be expanded simultaneously in momentum and frequency around $\omega=0$ as it can already be seen from perturbative calculations from ref.~\cite{Weldon_2001}, which corresponds in our truncation to small $\lambda$ and vanishing Landau damping $\gamma_k=0$.
We define the (Euclidean) vertex by the functional derivatives 
\begin{equation}
    \Gamma_k^{\mathrm{E}\,(4,0)}(x_1,\ldots x_4) = \frac{\delta^4 \Gamma_k[\phi; g]}{\delta \phi(x_1)\cdots \delta \phi(x_4)}\bigg|_{\phi=0},
    \label{eq:def_interaction_vertex}
\end{equation}
where the correlation function is evaluated on the Matsubara torus such that $x_j = (\tau_j, \mathbf{x}_j)$ with imaginary time $\tau_j$. 
The momentum space representation is introduced by 
\begin{equation}
    \begin{split}
        \Gamma_k^{\mathrm{E},(4,0)}&(x_1,\ldots, x_4) = \sum_{\omega_{n_1},\ldots, \omega_{n_4}} \int_{\mathbf{p}_1,\cdots \mathbf{p}_4} \\
        \times& \eu^{-\sum_{j=1}^4(\iu \omega_{n_j} \tau_j - \iu \mathbf{p}_j \mathbf{x}_j)} \quad \Gamma^{\mathrm{E},(4,0)}_k(p_1,\ldots p_4),
    \end{split}
\end{equation}
where $p_j=(\iu \omega_{n_j},\mathbf{p}_j)$ for $j\in\{1,\ldots 4\}$ with $\omega_{n_j}$ being a Matsubara frequency.
Since momentum is conserved at the vertex, we can write the fourpoint function as 
\begin{equation}
    \begin{split}
        \Gamma_k^{(4,0)}(p_1,\ldots, p_4) = \beta (2\pi)^3 \delta^{(3)}(p_1+\ldots+p_4)\times \\
        \delta_{0(n_1+\ldots +n_4)}V(p_1,p_2,p_3),
    \end{split}
\end{equation}
with the vertex function $V(p_1,p_2,p_3)$ and Kronecker delta $\delta_{mn}$.
Similarly to the propagator, knowledge of the vertex function at the Matsubara frequencies completely determines the analytic structure except for the lines where $\mathrm{Im}\, p_j^0 = 0$.
Different real-time correlation functions can be obtained by taking each frequency onto the real line with $\smash{\iu \omega_{n_j} \rightarrow p^0_j + \iu\epsilon_j}$, where $p^0_j\in\mathbb{R}$ and $\epsilon_j$ being infinitesimal but either smaller or greater zero.
There are in total 32 possibilities in choosing the infinitesimals which give rise to different correlation functions, because not only the $\epsilon$'s must be chosen but also the sum of each two of them, as discussed in refs.~\cite{BaymMermin,Evans:1990qh,Evans:1991ky}. 
To find the usual retarded/advanced correlators, one needs to choose one infinitesimal negative/positive while the others stay positive/negative. 
This also completely fixes the sum of all pairs, as the infinitesimals obey the condition 
\begin{equation}
    \sum_{j=1}^4 \epsilon_j = 0, 
\end{equation}
arising from conservation of momentum.

We define the sign of the infinitesimals and their sum of pairs by $s_j=\mathrm{sgn}(\epsilon_j)$ and $s_{ij} = \mathrm{sgn}(\epsilon_i + \epsilon_j)$.
One retarded correlator can now be given by the continuation $s_1=s_2=s_3=-s_4=+1$, which fixes the sign of the sums to $s_{12}=s_{13}=s_{23}=-s_{13}=-s_{34}=-s_{24}$.
Regarding the projection operator of the $\kappa_k$-vertex, we choose the latter, which is one of the eight retarded correlation functions. 
Thus, we can choose $\epsilon_1 = \epsilon_2 = \epsilon_3 = \epsilon$ and $\epsilon_4 = -3\epsilon$ and define one retarded vertex function as
\begin{equation}
    \begin{split}
        &V_k^{+++}(p_1,p_2,p_3) = \\
        &\quad\qquad\lim_{\epsilon\rightarrow 0}V_k^{\mathrm{E}}(p_1^0 + \iu\epsilon, p^0_2 +\iu\epsilon, p^0_3 + \iu\epsilon; \mathbf{p}_1,\mathbf{p}_2, \mathbf{p}_3),
    \end{split}
\end{equation}
with $p_i^0$ being real.
Of course, other retarded correlators can be built by choosing a different infinitesimal to be negative. 
Swapping all signs returns the advanced correlators on the other hand. 
Evaluating the vertex function explicitly in the complex plane, we find that 
\begin{equation}
    \begin{split}
        V_k^{+++}&(p_1, p_2, p_3) = \lambda_0 + \mathcal{A}(p^0_\mathrm{s}+2\iu\epsilon, \mathbf{p}_s) \\
        &+ \mathcal{A}(p^0_\mathrm{t}+2\iu\epsilon, \mathbf{p}_t) + \mathcal{A}(p^0_\mathrm{u}-2\iu\epsilon, \mathbf{p}_u),
    \end{split} 
    \label{eq:fourpoint_function}
\end{equation}
where we defined the combinations 
\begin{equation}
\begin{split}
    p_s =& (p^0_\mathrm{s},\mathbf{p}_s) = (p^0_1 + p^0_2, \mathbf{p}_1 + \mathbf{p}_2), \\
    p_t =& (p^0_\mathrm{t},\mathbf{p}_t) = (p^0_1 + p^0_3, \mathbf{p}_1 + \mathbf{p}_3), \\
    p_u =& (p^0_\mathrm{u},\mathbf{p}_u) = (p^0_1 + p^0_4, \mathbf{p}_1 + \mathbf{p}_4),
\end{split}
\end{equation} 
and we also define
\begin{equation}
    \mathcal{A}(z,\mathbf{p}) = \iu z \mathrm{s}_\mathrm{I}(z)\kappa_k(\mathbf{p}^2).
    \label{eq:def_mathcalA}
\end{equation}
Evaluating this specific continuation given above we find that 
\begin{equation}
    \begin{split}
        V_k^{+++}(p_\mathrm{s},p_\mathrm{t},p_\mathrm{u}) = \lambda_0 &+ \iu p^0_\mathrm{s} \kappa_k(\mathbf{p}_s^2) + \iu p^0_\mathrm{t} \kappa_k(\mathbf{p}_t^2) \\
        &- \iu p^0_\mathrm{u} \kappa_k(\mathbf{p}_u^2).
    \end{split}
    \label{eq:vertex_function_retarded}
\end{equation}
We may represent the fourpoint function via the diagrams 
\begin{equation}
    \fourvertexNOLABEL{0.25} = \fourvertexpointlikeNOLABEL{0.25} + \fourvertexsignoperatorNOLABEL{0.25} ,
\end{equation}
where the first diagram on the right-hand side represents the $\lambda$ interaction and the second the $\mathcal{A}$ insertion, proportional to the sign-operator.
This damping term captures the imaginary part of the spacelike 2-by-2 scattering amplitude.
Thus, it should be regarded as being defined only for spacelike momenta $\smash{p^2=-(p^0)^2+\mathbf{p}^2>0}$ and the vertex part $\mathcal{A}(z,\mathbf{p})$ vanishes for time- or light-like momenta $p^2\leq 0$. 
Furthermore, the function $\mathcal{A}(z,\mathbf{p})$ given in \cref{eq:def_mathcalA} vanishes exactly at $z=0$. 
Using these assumptions and the explicit expression for the vertex in momentum space we will set $\mathcal{A}(0,\mathbf{p})=0$ for any $\mathbf{p}$.

The explicit form \cref{eq:vertex_function_retarded} suggests that a projector for the coupling $\kappa_k$ can be built with 
\begin{equation}
    \begin{split}
        \kappa_k(\mathbf{p}_s^2) = \frac{\partial }{\partial p^0_\mathrm{s} } \mathrm{Im}V^{+++}_k(p_\mathrm{s},p_\mathrm{t},p_\mathrm{u})\bigg|_{p_\mathrm{u}=p_\mathrm{t}=0,\,p^0_\mathrm{s} = 0}.
    \end{split}
    \label{eq:def_projector_kappa}
\end{equation}

The remaining vertex that appears in the flow equations consist of both derivatives with respect to scalar fields and external metric fields.
The only relevant vertex that involves the external metric field is 
\begin{equation}
    \begin{split}
        \Gamma_k^{(2,1)\mu\nu}(x_1, x_2; y) =& \frac{\delta^3\Gamma_k[\phi; g]}{\delta \phi(x_1)\delta\phi(x_2)\delta g_{\mu\nu}(y)}\bigg|_{\phi=0}, \\
    \end{split}
\end{equation}
again evaluated on the Matsubara torus. 
We define the Fourier representation analogously to the previous ones.
Explicitly for the spacelike off-diagonal elements $(\mu,\nu) = (x,y)$ we define
\begin{equation}
    \begin{split}
        &\Gamma_k^{(2,1)xy}(x_1, x_2; y) = \\
        &\sum_{\omega_m, \omega_{n_1}, \omega_{n_2}}\int_{\mathbf{p}_1,\mathbf{p}_2,\mathbf{k}} \mkern-18mu\mkern-18mu\eu^{-\iu \omega_m y_0 +\iu \mathbf{k}\mathbf{y} - \sum_{i=1,2} (\iu\omega_{n_i} (x_i)_0 - \iu \mathbf{p}_i\mathbf{x}_i)} \\
        &\times \Gamma_k^{(2,1)xy}(p_1, p_2; k), \\
    \end{split}
\end{equation}
where
\begin{equation}
    \begin{split}
        \Gamma_k^{(2,1)xy}(p_1, p_2; k) = -\beta(2\pi)^3\delta^{(3)}(\mathbf{k}+\mathbf{p}_1+\mathbf{p}_2)\\
        \times\delta_{0(m+n_1+n_2)}p_1^x p_2^y.
    \end{split}
\end{equation}
The vertex can be continued onto the complex plane analogously to all previously encountered correlation functions.

\subsection{Initial Conditions}\label{ch:subsection_initial_conditions}
The initial conditions for the RG flow are chosen such that the discontinuous contributions vanish at $k=\Lambda$, thus $\Gamma_{\Lambda}^\text{d}[\phi; g]=0$.
With this choice the dissipative terms are being generated dynamically during the flow while also staying consistent with the microscopic formulation defined by \cref{eq:phi4_microscopic}, which does not contain any dissipative terms. 
The initial conditions are explicitly chosen such that
\begin{equation}
    \begin{split}
        m_\Lambda &= m,\\
        \gamma_\Lambda &= 0,\\
        \kappa_\Lambda\left(-\boldsymbol{\nabla}\,^2\right) &= 0,\\
        \eta_\Lambda &= \eta_\mathrm{UV}(\Lambda),
    \end{split}
    \label{eq:def_initial_conditions}
\end{equation}
where $m$ is the vacuum mass and $\eta_\text{UV}(\Lambda)$ is a function that depends on the UV cutoff of the RG scales such that formally $\smash{\eta_\text{UV}(\Lambda)\longrightarrow\infty}$ for $\Lambda\rightarrow\infty$.
The latter can be motivated by phenomenology and by studying the flow equations in the limit of high energy scales $k\gg m_k$.
In the perturbative regime, it was shown that the shear viscosity coefficient diverges as $\gamma_k^{-1}$ for small widths and is ill-defined in the limit $\gamma_k=0$ unless the scalar field is massless~\cite{Jeon_1992}.
Hence, there is no possibility for the shear viscosity to be well-defined if $k=\Lambda$ and $\gamma_k=0$, since the involved processes that produce the coefficient need to be at least of two-loop order. 
This demands a momentum dependent vertex in the flow equations, which is not present at the ultraviolet where $\kappa_k(\mathbf{p}^2)=0$.
While this behaviour is circumvented in perturbative computations by starting at the two-loop order, where decays into the heat bath can appear, we cannot do this in the continuous RG scheme unless we introduce a finite damping at the starting point $k=\Lambda$.
Since we want to generate the damping dynamically, starting at zero, we have to deal with the divergent $\eta_k$ flow. 
This problem is circumvented by introducing a regularized viscosity coefficient
\begin{equation} 
    \tilde{\eta}_k=\gamma_k \eta_k,
\end{equation} 
which has a well-defined flow, even at $k=\Lambda\rightarrow\infty$, and we set the initial condition to
\begin{equation}
    \tilde{\eta}_{\Lambda} = 0.
\end{equation} 
The viscosity coefficient can be reconstructed from the Landau damping and the regularized viscosity afterwards.
An analysis of the asymptotics of the flow equation of $\eta_k$ near $k=\Lambda$ will be given in \cref{ch:flowequations_asymptotics}.

\section{Flow Equations}
\label{ch:flowequations}
After introducing the FRG flow equation together with the truncation defined in \cref{eq:def_gamma_reg,eq:def_gamma_dis}, we will have a look at the flow equations of the couplings $m_k$, $\kappa_k(\mathbf{p}^2)$, $\gamma_k$ and $\eta_k$, defined by the projections in \cref{eq:def_projector_gammak,eq:def_projector_kappa,eq:eta_k_def}.
Before giving the exact relations, we will discuss how the Matsubara sums are generally calculated and how the analytic continuation is performed. 
We will show explicitly how this can be done for the imaginary part of the tadpole diagram, which appears in the Landau damping flow equation.

\subsection{Matsubara Sums \& Analytic Continuation}
The specific form of the discontinuous parts introduced in \cref{eq:def_gamma_dis} allows us to analytically determine all Matsubara sums that appear in any flow equation. 
In general, the sum can be calculated once the analytic structure of the corresponding diagram is known for complex valued frequencies. 
Isolated poles and branch cuts might be introduced by the propagator $G_k(z,\mathbf{p})$ or the discontinuous vertex $\mathcal{A}(z,\mathbf{p})$.
In both cases, the Matsubara sum can be calculated analytically and expressed by some threshold function, once the external frequencies have been continued onto the real line. 
We will now explain how any Matsubara sum appearing in the flow equations can be calculated. 
For some guidance, we will explicitly show this on the sum appearing in the imaginary part of the tadpole diagram, which is part of the Landau damping flow defined by the projection \eqref{eq:def_projector_gammak}.
The arguments presented for this example generalize to any other Matsubara sum encountered in this work. 

\begin{figure}[htbp]
    \centering
    \includegraphics[width=0.99\columnwidth]{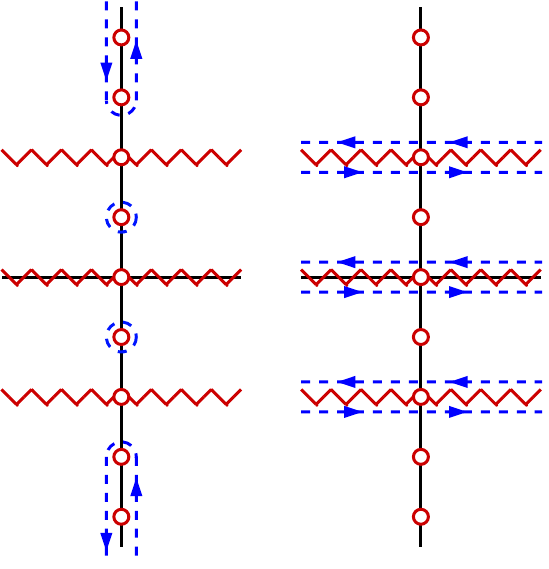}
    \caption{Analytic structure of the tadpole diagram at $\gamma_k>0$ in the complex frequency plane.
    The red lines indicate branch cuts, while isolated poles from the Bose distribution are represented by red circles. 
    The blue paths show the integration contour $\mathcal{C}$ in the left panel, corresponding to the Matsubara sum \eqref{eq:tadpole_matsubara_sum_contour_matsubara_poles}. 
    The modes at zero, as well as $\pm \iu \omega_m$ are excluded due to the presence of a branch cut.
    An equivalent contour corresponding to \cref{eq:sum_tadpole_intermediate} is shown in the right panel, enclosing only the branch cuts.}
    \label{fig:contour_tadpole}
\end{figure}

First, recall that the flow equations are all one-loop, and thus we only have to consider a single Matsubara sum for each loop integral over the internal Matsubara frequencies $\omega_n$. 
The diagrams can of course depend on external spatial momenta $\mathbf{p}_i$ and external Matsubara frequencies $\smash{\omega_{m_i}}$ and the spatial loop momentum will be labelled as $\mathbf{q}$.
When considering the tadpole diagram with only an insertion of the discontinuous vertex $\mathcal{A}(z,\mathbf{p})$, represented by the diagram
\begin{equation*}
    \selfenergysignoperatorannotations{0.5},
\end{equation*}
the corresponding frequency sum is given by 
\begin{equation}
    \begin{split}
        \mathcal{M}_\mathcal{A}(\iu \omega_m, \mathbf{p}; \mathbf{q}) = \sum_{\omega_n} f(\iu\omega_n),
    \end{split}
    \label{eq:tadpole_matsubara_sum}
\end{equation}
with $f(z)$ being the product of the analytically continued propagator $G_k(z,\mathbf{p})$ and discontinuous vertex $\mathcal{A}(z,\mathbf{p})$, such that 
\begin{equation}
    \begin{split}
        f(z) = G_k(z,\mathbf{q})\big[ &\mathcal{A}(0,\mathbf{0}) + \mathcal{A}(z + \iu \omega_m,\mathbf{q} + \mathbf{p})\\
        + &\mathcal{A}(z - \iu \omega_m,\mathbf{q} - \mathbf{p}) \big]\\
        = G_k(z,\mathbf{q})\big[ &\mathcal{A}(z + \iu \omega_m,\mathbf{q} + \mathbf{p})\\
        + &\mathcal{A}(z - \iu \omega_m,\mathbf{q} - \mathbf{p})\big],
    \end{split}
\end{equation}
where we used that $\mathcal{A}(0,\mathbf{0})=0$.
Note that we suppress the explicit dependence of $f$ on the external momentum and frequency and internal spatial momentum for brevity.
The function $f(z)$ exhibits branch cuts from the $\mathcal{A}(z,\mathbf{p})$ insertion parallel to the real line at $z=\pm \iu \omega_m$, where the external Matsubara frequency is located. 
Additionally, the propagator introduces a branch cut on the real line, either over the complete frequency range $(-\infty,\infty)$ if $\gamma_k>0$ or for real valued frequencies which satisfy $\smash{|\omega|\ge \sqrt{m_k^2+k^2+\mathbf{q}^2}}$ if $\gamma_k=0$.
The analytic structure is shown in \cref{fig:contour_tadpole} at a finite damping $\gamma_k>0$.

The sum \eqref{eq:tadpole_matsubara_sum} can be expressed by a sum of integrals along the discontinuities, similar to the treatment in refs.~\cite{Heckmann_2011,Tripolt_2015}.
To achieve this, it is beneficial to exclude the modes from the sum which are located on top of a cut. 
By the use of Cauchy's theorem, we can express the sum as the contour integral 
\begin{equation}
    \begin{split}
        \sum_{\omega_n} f(\iu\omega_n) =& \int_\mathcal{C} (1 + 2 \mathrm{n}_\mathrm{B}(z))f(z) \frac{\dd z }{4\pi \iu }\\
        &+Tf(0) + Tf(\iu \omega_m) + T f(-\iu \omega_m),
    \end{split}
    \label{eq:tadpole_matsubara_sum_contour_matsubara_poles}
\end{equation}
where 
\begin{equation}
    \mathrm{n}_\mathrm{B}(z) = \frac{1}{\eu^{\beta z }-1},
\end{equation}
is the Bose-distribution.
The contour $\mathcal{C}$ is depicted in \cref{fig:contour_tadpole}, enclosing all Matsubara modes except $0$, $\smash{\omega_m}$ and $\smash{-\omega_m}$, as they are located on a branch cut of either the propagator or the vertex function.
Other diagrams may contain more branch cuts which run through the poles of the Matsubara frequencies. 
For each cut, the corresponding Matsubara pole is excluded from the contour integral and added explicitly afterwards.

Since the function $f(z)$ does not contain any further isolated poles or branch cuts, the contour $\mathcal{C}$ can now be deformed to several line integrals above and below each cut, see \cref{fig:contour_tadpole}, which corresponds to the integrals 
\begin{equation}
    \begin{split}
        \sum_{\omega_n}&f(\iu\omega_n) = Tf(0) + Tf(\iu \omega_m) + T f(-\iu \omega_m) \\
        +&\sum_{a\in\{0,\pm\iu \omega_m\}}\int_{-\infty + a + \iu\epsilon}^{\infty + a + \iu\epsilon} \left( 1+2\mathrm{n}_\mathrm{B}(z) \right)f(z)\frac{\dd z }{4\pi \iu }\\
        +&\sum_{a\in\{0,\pm\iu \omega_m\}}\int_{\infty + a - \iu\epsilon}^{-\infty + a - \iu\epsilon} \left( 1+2\mathrm{n}_\mathrm{B}(z) \right)f(z)\frac{\dd z }{4\pi \iu }.\\
    \end{split}
    \label{eq:sum_tadpole_intermediate}
\end{equation}
The last two terms can be simplified by a change of variables to integrals along the discontinuities of the integrand.
The latter is defined for a function $f(x)$ and $x\in\mathbb{R}$ by 
\begin{equation}
    \mathrm{Disc}[f](x) = \lim_{\epsilon\rightarrow 0} \frac{\iu }{2}\left[ f(x+\iu\epsilon) - f(x-\iu\epsilon)  \right],
    \label{eq:def_disc}
\end{equation}
also found in ref~\cite{LeBellac_1996}.
With this definition, we can express \cref{eq:sum_tadpole_intermediate} as 
\begin{equation}
    \begin{split}
        \sum_{\omega_n}&f(\iu\omega_n) = Tf(0) + Tf(\iu \omega_m) + T f(-\iu \omega_m) \\
        -&\sum_{a\in\{0,\pm\iu \omega_m\}}\int_{-\infty}^{\infty} \mathrm{Disc}[\left( 1+2\mathrm{n}_\mathrm{B}\right)\cdot f](z + a)\frac{\dd z }{2\pi }.\\
    \end{split}
    \label{eq:sum_tadpole_integral_along_disc}
\end{equation}
The discontinuity of the integrand is determined by the branch cuts of the propagator and vertices, as well as the singularity of the Bose distribution at the Matsubara frequencies.
Since the latter is only present at the Matsubara frequencies and the branch cut of the function $f(z)$ vanishes exactly on the imaginary axis, since $\mathcal{A}(0,\mathbf{q})=0$ and $\smash{G_k(0, \mathbf{q})=1/(m_k^2+k^2+\mathbf{q}^2)}$.
This allows us to express the discontinuity as 
\begin{equation}
    \begin{split}
        \mathrm{Disc}[1+2\mathrm{n}_\mathrm{B}\cdot f](z + a) =& (1+2\mathrm{n}_\mathrm{B}(z))\mathrm{Disc}[f](z+a)\\
        &+ 2\mathrm{Disc}[\mathrm{n}_\mathrm{B}](z)f(z+a),
    \end{split}
    \label{eq:tadpole_disc_decomposition}
\end{equation}
where we have used the periodicity of the Bose-distribution on the Matsubara modes where $\mathrm{n}_\mathrm{B}(z + \iu \omega_m) = \mathrm{n}_\mathrm{B}(z)$.

The discontinuity corresponds to an isolated singularity at the origin. 
To calculate this quantity, we expand around small $\beta(z\pm \iu\epsilon)$, such that 
\begin{equation}
    \begin{split}
        \frac{1}{\eu^{\beta(z\pm \iu \epsilon)}-1} \approx \beta^{-1} \frac{1}{z \pm \iu \epsilon},
    \end{split}
\end{equation}
and use the identity
\begin{equation}
    \frac{1}{z \mp \iu\epsilon} \rightarrow \mathcal{P}\left( \frac{1}{z } \right) \pm \iu\pi\delta(z),
\end{equation}
with $\mathcal{P}$ denoting the Cauchy principal value.
Combining the last two equations with the definition of the discontinuity \eqref{eq:def_disc}, we see that 
\begin{equation}
    \mathrm{Disc}[\mathrm{n}_\mathrm{B}](z) = \frac{\pi }{\beta }\delta(z).
\end{equation}
Thus, the second term in \cref{eq:tadpole_disc_decomposition} cancels the previously excluded Matsubara modes at $0$ and $\pm \iu \omega_m$.
The sum is therefore completely determined by the integral 
\begin{equation}
    \begin{split}
        &\sum_{\omega_n}f(\iu\omega_n) = \\
        &-\sum_{a\in\{0,\pm\iu \omega_m\}} \int_{-\infty}^\infty (1+2\mathrm{n}_\mathrm{B}(z))\mathrm{Disc}[f](z+a)\frac{\dd z }{2\pi },
    \end{split}
    \label{eq:tadpole_matsubara_sum_final_with_disc}
\end{equation}
with 
\begin{equation}
    \begin{split}
        \mathrm{Disc}[f](z) =& -\frac{1}{2} \rho_k(z, \mathbf{q})[\mathcal{A}(z + \iu \omega_m, \mathbf{q} + \mathbf{p}) \\
        &+ \mathcal{A}(z - \iu \omega_m, \mathbf{q} - \mathbf{p})],\\
        \mathrm{Disc}[f](z\pm \iu \omega_m) =& -z G_k(z\pm \iu \omega_m, \mathbf{q})\kappa_k((\mathbf{q} \mp \mathbf{p})^2).
    \end{split}
    \label{eq:tadpole_mastubara_sum_euclidean}
\end{equation}
Any other sum encountered in the flow equations can be brought into the form \eqref{eq:tadpole_matsubara_sum_final_with_disc}.

Since the frequency sum is formally carried out in \cref{eq:tadpole_mastubara_sum_euclidean}, we can now continue the external Matsubara frequency to the real line. 
Specifically for the Landau damping flow, we are interested in the imaginary part of the retarded two-point function, as defined in \cref{eq:def_projector_gammak}.
Choosing the retarded contour with the continuation $\iu \omega_m \rightarrow \omega + \iu\epsilon$, and taking the limit $\epsilon\rightarrow 0$, we find for the imaginary part of the sum 
\begin{equation}
    \begin{split}
        \mathrm{Im}\,&\mathcal{M}^\mathrm{R}_\mathcal{A}(\omega, \mathbf{p}; \mathbf{q}) = \int_{-\infty}^\infty \frac{\dd z }{4\pi } (1+2\mathrm{n}_\mathrm{B}(z))\bigg[\rho_k(z,\mathbf{q})\\
        \times&[(z + \omega)\kappa_k((\mathbf{q} + \mathbf{p})^2) + (\omega - z )\kappa_k((\mathbf{q} - \mathbf{p})^2)]\\
        +& 2z G_k^\mathrm{R}(z + \omega,\mathbf{q})\kappa_k((\mathbf{q} - \mathbf{p})^2) \\
        +& 2z G_k^\mathrm{A}(z - \omega,\mathbf{q})\kappa_k((\mathbf{q} + \mathbf{p})^2)\bigg],
    \end{split}
\end{equation}
with $G_k^\mathrm{A}$ being the advanced propagator.
Evaluating this quantity at vanishing external momentum and at $\omega=m_k$ yields 
\begin{equation}
    \begin{split}
        \mathrm{Im}\,\mathcal{M}^\mathrm{R}_\mathcal{A}&(m_k, \mathbf{0}; \mathbf{q}) = \int_{-\infty}^\infty \frac{\dd z}{4\pi} (1+2\mathrm{n}_\mathrm{B}(z))\kappa_k(\mathbf{q}^2)\\
        \times&\bigg[2 m_k\rho_k(z,\mathbf{q}) + z [\rho_k(z + m_k,\mathbf{q}) - \rho_k(z - m_k,\mathbf{q})]\bigg]. \\
    \end{split}
\end{equation}
If the damping is vanishing, the integral can be performed straightforwardly since $\rho(z,\mathbf{q})$ is given by a Dirac delta.
In the case where $\gamma_k>0$, the expression can be identified within a general class of integrals which appear in all flow equations in this work.
These are of the form 
\begin{equation}
    \begin{split}
        \mathcal{I}_m^n&(|\mathbf{p}_1|,\ldots,|\mathbf{p}_m|; y) = \int_{-\infty}^{\infty}\!\frac{(1+2\mathrm{n}_\mathrm{B}(z)) z^n }{\prod_{j=1}^m\prod_{l=1}^4(z+y-z_l^{(j)})}\dd z,
    \end{split}
    \label{eq:def_threshold_function}
\end{equation}
with $m,n$ being positive, natural numbers and $y$ real. 
The quantity $z_l^{(j)}$ is the $l$-th pole of the spectral density \eqref{eq:spectral_function_truncation} evaluated at the momentum $\mathbf{p}_j$.
The function $\mathcal{I}_m^n$ can be decomposed into a linear combination of polygamma functions $\psi^{(n)}(z)$ and its general form can be found in \cref{ch:appendix_thresholdfunctions}.
For the tadpole diagram this corresponds to the expression 
\begin{equation}
    \begin{split}
        &\mathrm{Im}\, \mathcal{M}_\mathcal{A}^\mathrm{R}(m_k, \mathbf{0}; \mathbf{q}) = \frac{\gamma_k \kappa_k(\mathbf{q}^2)}{2\pi}\bigg[\\
        &\quad\mathcal{I}_1^2(|\mathbf{q}|; m_k)-\mathcal{I}_1^2(|\mathbf{q}|; -m_k)\\
        &+m_k\left( \mathcal{I}_1^1(|\mathbf{q}|; 0) + \mathcal{I}_1^1(|\mathbf{q}|; m_k) + \mathcal{I}_1^1(|\mathbf{q}|; -m_k) \right)\bigg].
    \end{split}
\end{equation}
Furthermore, if the Landau damping is considered small, such that the poles of the spectral density can be described by $z=\pm \Omega_q \pm \iu\gamma_k/2$, the expression above simplifies considerably to 
\begin{equation}
    \begin{split}
    &\mathrm{Im}\, \mathcal{M}_\mathcal{A}^\mathrm{R}(m_k, 0; \mathbf{q}) = - \frac{8\pi \kappa_k(\mathbf{q}^2)}{\beta(\gamma_k^2 + 4\Omega_q^2)} \\
    &+\frac{\kappa_k(\mathbf{q}^2)}{4m_k\Omega_q }\bigg[ 4\iu m_k \psi^{(0)}\left( \frac{\beta(\gamma_k - 2 \iu \Omega_q)}{4\pi } \right) \\
    &(2\iu (m_k - \Omega_q) - \gamma_k )\psi^{(0)}\left( \frac{\beta(\gamma_k - 2\iu (m_k - \Omega_q))}{4\pi }  \right) \\
    &(\gamma_k - 2\iu (m_k + \Omega_q))\psi^{(0)}\left( \frac{\beta(\gamma_k - 2\iu (m_k + \Omega_q))}{4\pi }  \right) + \text{c.c.}\bigg],
    \end{split}
\end{equation}
with c.c. referring to the complex conjugate
The other Matsubara sums appearing in this work can similarly be expressed by the function $\mathcal{I}_m^n$. 
Explicit calculations can be found in \cref{ch:appendix_contour_integrals}. 

\subsection{Flow Equations}
We can now give the complete set of flow equations for the four coefficients $\smash{m_k, \gamma_k, \kappa_k(\mathbf{p}^2)}$ and $\eta_k$. 
A detailed derivation of each contour integral can be found in \cref{ch:appendix_contour_integrals}.
By acting with the projectors defined in \cref{eq:def_projector_gammak,eq:def_projector_kappa,eq:def_projector_mk,eq:eta_k_def} on the Wetterich equation \eqref{eq:def_wetterich_equation}, we find the following set of differential equations
\begin{equation}
    \begin{split}
        &\partial_t m_k^2 =\frac{1}{2^4\pi^3 } \frac{\tilde{\partial}}{\tilde{\partial} t} \int_0^\infty \dd q  q^2 \bigg[  \gamma_k\lambda \mathcal{I}_1^1(q; 0) \\
        & \,\,+ 8\kappa_k(q^2)\bigg((m_k^2+k^2+q^2)\mathcal{I}_1^1(q; 0) -\mathcal{I}_1^3(q; 0) \bigg)\bigg],  \\
    \end{split}
    \label{eq:flow_mass}
\end{equation}
\begin{equation}
    \begin{split}
    \partial_t \gamma_k =& \frac{\gamma_k}{2^4 \pi^3 m_k}\frac{\tilde\partial }{\tilde\partial t}\int_0^\infty q^2 \kappa_k(q^2) \bigg[\mathcal{I}_1^2(|\mathbf{q}|; m_k) \\
    &- \mathcal{I}_1^2(|\mathbf{q}|; -m_k) + m_k\big( \mathcal{I}_1^1(|\mathbf{q}|; 0) \\
    &+ \mathcal{I}_1^1(|\mathbf{q}|; m_k) + \mathcal{I}_1^1(|\mathbf{q}|; -m_k) \big)\bigg] \,\dd q ,
    \end{split}
    \label{eq:flow_gamma}
\end{equation}
\begin{equation}
    \begin{split}
    \mathbf{p}^2\partial_t \kappa_k(\mathbf{p}^2) =& -\mathbf{p}^2\frac{\lambda^2 \gamma_k^2}{2^6\pi^3}\frac{\tilde \partial }{\tilde \partial t } \int_{-1}^1 \dd \cos\theta \int_0^\infty \dd q \, q^2 \\ 
    &\times  \beta\partial_\beta \mathcal{I}_2^1(q, |\mathbf{q}+\mathbf{p}|; 0) + \mathcal{O}(\lambda \, \kappa_k),\\
    \end{split}
    \label{eq:flow_kappa}
\end{equation}
\begin{equation}
    \begin{split}
    \partial_t \eta_k =& \frac{\gamma_k^2 }{160 \pi^3}\frac{\tilde{\partial}}{\tilde{\partial} t} \beta\frac{\partial }{\partial \beta} \int_0^\infty \mathcal{I}_2^1(q,q; 0)q^6\dd q, \\
\end{split}
    \label{eq:flow_eta}
\end{equation}
\begin{equation}
    \begin{split}
        \partial_t \tilde\eta _k =& \gamma_k \partial_t \eta_k + \eta_k \partial_t \gamma_k.
    \end{split}
    \label{eq:flow_eta_tilde}
\end{equation}
The symbol $\tilde\partial/(\tilde\partial t )$ is the \textit{formal derivative}, defined by 
\begin{equation}
    \frac{\tilde \partial }{\tilde \partial t} = \frac{\partial R_k}{\partial t} \frac{\partial }{\partial R_k},
\end{equation}
such that it only acts on the regulator.
Note that the flow equation of $\kappa_k$ is truncated at order $\kappa_k$, as this coupling is much smaller than the quartic coupling $\lambda$ and even for $\lambda~\sim\mathcal{O}(1)$ the next order term is only relevant in the high-temperature regime.
Furthermore, we solve the quantity $\mathbf{p}^2\kappa_k(\mathbf{p}^2)$ instead of just $\kappa_k(\mathbf{p}^2)$, due to a divergence appearing at $\gamma_k=0$ in the vertex damping at vanishing external momentum.
This can already be seen from perturbative results, see for example ref~\cite{Weldon_1983}, where it was shown that the imaginary part of the diagram 
\begin{equation*}
    \bubble{0.5}
\end{equation*}
is non-analytic in the origin where $\omega=0$ and $\mathbf{p}=0$.
\begin{figure}[htbp]
    \includegraphics[width=0.60\columnwidth]{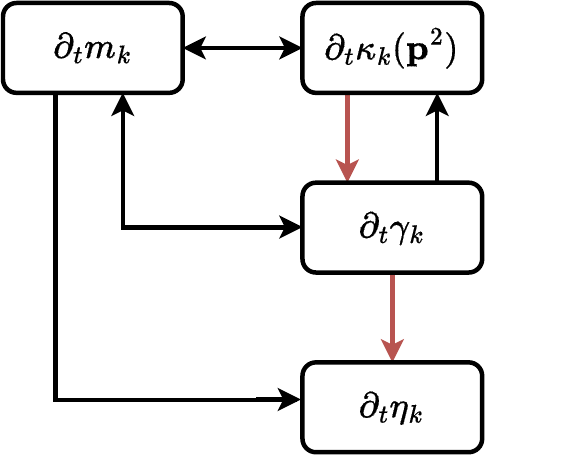}
    \caption{Dependence between the flow equations.
                A red arrow indicates a necessary existence of the prerequisite.}
    \label{fig:sketch_flow_dependences}
\end{figure} 
Any momentum integral in which the coupling $\kappa_k$ appears is however well-defined as the divergence is of order $\smash{|\mathbf{p}|^{-1}}$, which is either cancelled by the integral measure of the momentum integrals or regularized by a finite external momentum.
Also, we will restrict the flow of the mass $m_k$ in the following results to thermal contributions only and neglect any contributions to the vacuum part. 
A full treatment of vacuum flow will be reserved for future work. 

The coupled system of equations effectively splits into two parts. 
The flows of the mass, Landau damping and the vertex damping all couple to each other and cannot be solved independently. 
While the mass flows already at vanishing damping, the flow of $\gamma_k$ is vanishing if $\kappa_k(\mathbf{p}^2)=0$ as mentioned in \cref{ch:subsection_initial_conditions}. 
Once a finite imaginary part of vertex is generated, the Landau damping starts to develop. 
The flow of the shear viscosity, and the regularized viscosity, do not feed back into the flow equations of the other coefficients but depend on all three.
This hierarchy is shown in figure \cref{fig:sketch_flow_dependences}.

\subsection{Asymptotics of the viscosity flow}\label{ch:flowequations_asymptotics}
As it was mentioned in \cref{ch:subsection_initial_conditions}, the viscosity coefficient is ill-defined at $k=\Lambda$ when there is no thermal broadening of the mass pole and the spectral density is given by a Dirac delta.
In this limit, not only the coefficient but additionally the flow $\partial_t \eta_k$ is ill-defined.
This can be seen by expanding \cref{eq:flow_eta} around small $\gamma_k$, which is approximating the flow close to the initial condition. 
Using 
\begin{equation}
    \begin{split}
        \gamma_k^2 \mathcal{I}_2^1(q,q) = \gamma_k^{-1} \frac{\pi\beta }{2^5(m_k^2 + q^2 +k^2)}\mathrm{n}_\mathrm{B}\left(\sqrt{m_k^2 + q^2+k^2}\right) \\
        \times
        \left(1+\mathrm{n}_\mathrm{B}\left(\sqrt{m_k^2 + q^2+k^2}\right)\right) + \mathcal{O}(\gamma_k),
    \end{split}
    \label{eq:I21_asymptotics}
\end{equation}
we see that the flow of $\eta_k$ is expected to diverge as $\gamma_k\rightarrow 0$. 
Using the other flow equations, one can analytically determine the Landau damping and vertex-damping coefficients at large scales $k\approx \Lambda$ to be given by 
\begin{equation}
    \begin{split}
        \kappa_k(\mathbf{p}^2)|_{k\approx \Lambda} &= -\frac{1}{16\pi \sqrt{\mathbf{p}^2}}\mathrm{n}_\mathrm{B}\left(\frac{1}{2}\sqrt{4(m^2+k^2)+\mathbf{p}^2}\right), \\
        \gamma_k|_{k\approx \Lambda} &= \frac{k \sinh(\beta m )}{2^7\pi^3 \beta m }\eu^{-2\beta k },
    \end{split}
    \label{eq:kappa_gamma_asymptotics}
\end{equation} 
where the first equation can be found from eq. (C2) in ref~\cite{Weldon_2001} by expanding around a vanishing frequency and a derivation of the Landau damping asymptotics is outlined in \cref{ch:appendix_asymptotics}.
Equation \eqref{eq:I21_asymptotics} and the asymptotic scaling of the Landau damping show that $\partial_t \eta_k\sim \eu^{\beta k}$ in the UV, which makes it difficult to define the coefficient, and the flow equation itself at the microscopic scale. 
However, the regularized viscosity coefficient is defined such that the exponential contributions $\eu^{\beta k}$ cancel out. 
In fact, when expanding the flow equation \eqref{eq:flow_eta_tilde} around $\gamma_k=0$, we find that the flow of $\tilde \eta_k$ is governed at asymptotically large $\beta k$ by 
\begin{equation}
    \partial_t \tilde\eta_k|^{k\approx\Lambda_\mathrm{UV}} = \sqrt{\frac{9}{32\pi}}\frac{m \beta^{-3}\eu^{-\beta\sqrt{m^2+k^2}}}{(\beta\sqrt{m^2+k^2})^{3/2}\sinh(\beta m)}.
    \label{eq:eta_tilde_asymptotics}
\end{equation}
The equation is (numerically) stable even at the initial condition.

\subsection{Solutions of the Flow Equations}
We will now show some solutions to the flow equations defined by \cref{eq:flow_eta,eq:flow_eta_tilde,eq:flow_gamma,eq:flow_kappa,eq:flow_mass}.
In the following, we set the quartic coupling $\smash{\lambda=10^{-2}}$ while varying the temperature. 
The UV cutoff was set to $\smash{\Lambda=120 m}$.
To solve the flows, the momentum dependence of $\kappa_k$ needs to be resolved on the full domain $\smash{0\leq |\mathbf{p}|\leq \Lambda}$. 
This is done by solving the flow \cref{eq:flow_kappa} on a momentum grid of 100 Chebychev nodes. 
\begin{figure}[htbp]
    \centering 
    \includegraphics[width=1.0\columnwidth]{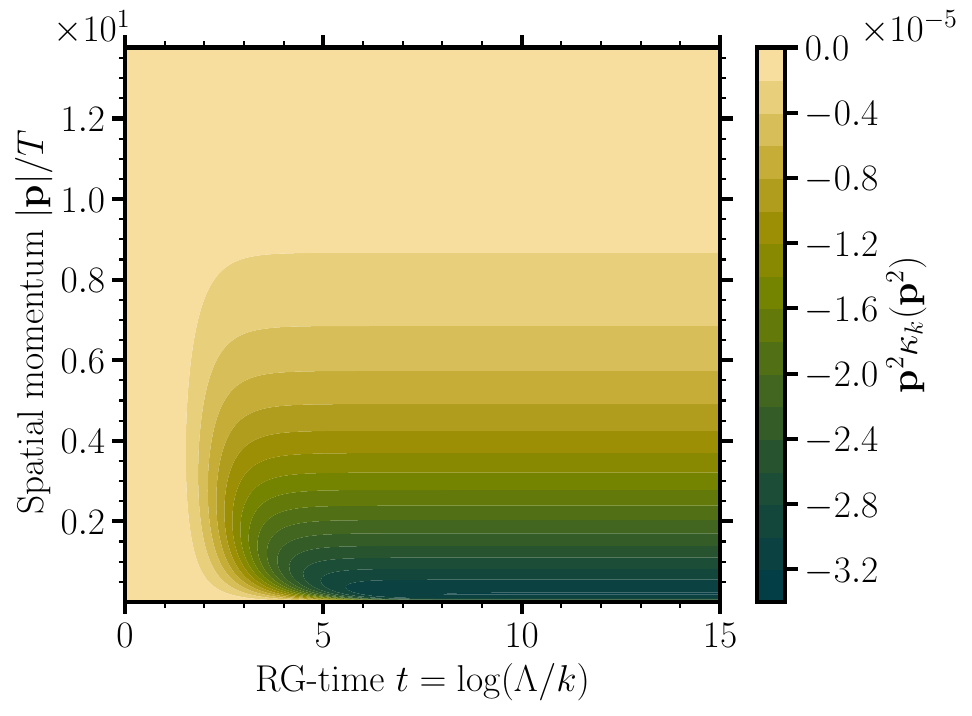}
    \caption{Solution of the four vertex flow $|\mathbf{p}|^2\kappa(\mathbf{p}^2)$ at fixed temperature $T=m$ and coupling $\lambda=10^{-2}$.} 
    \label{fig:solution_kappa_flow_alternative}
\end{figure} 
From the solutions on the momentum grid, an interpolating function is constructed using a cubic-spline, which is used in flows \eqref{eq:flow_gamma} and \eqref{eq:flow_mass}.
The solution to the flow equation \cref{eq:flow_kappa} is shown in \cref{fig:solution_kappa_flow_alternative}.
At scales below the temperature, the coefficients remain approximately zero.
The flow starts to deviate from zero when $k$ is close to the temperature $T$.
The coupling $\kappa_k$ develops a peak at small momenta while decaying exponentially for momenta large compared to the temperature.
Note that only the contributions $\sim\lambda^2$ are considered here, since $\kappa$ is already small and thus contributions $\sim \lambda \kappa_k$ and $\sim\kappa_k^2$ are negligible. 
The latter two contributions are only relevant in the region where the quartic interaction and the temperature are large as the coefficient $\kappa_k$ is strongly related to thermal effects of the $\phi^2$-$\phi^2$ scattering process.

\begin{figure}[htbp]
    \centering
    \includegraphics[width=0.9\columnwidth]{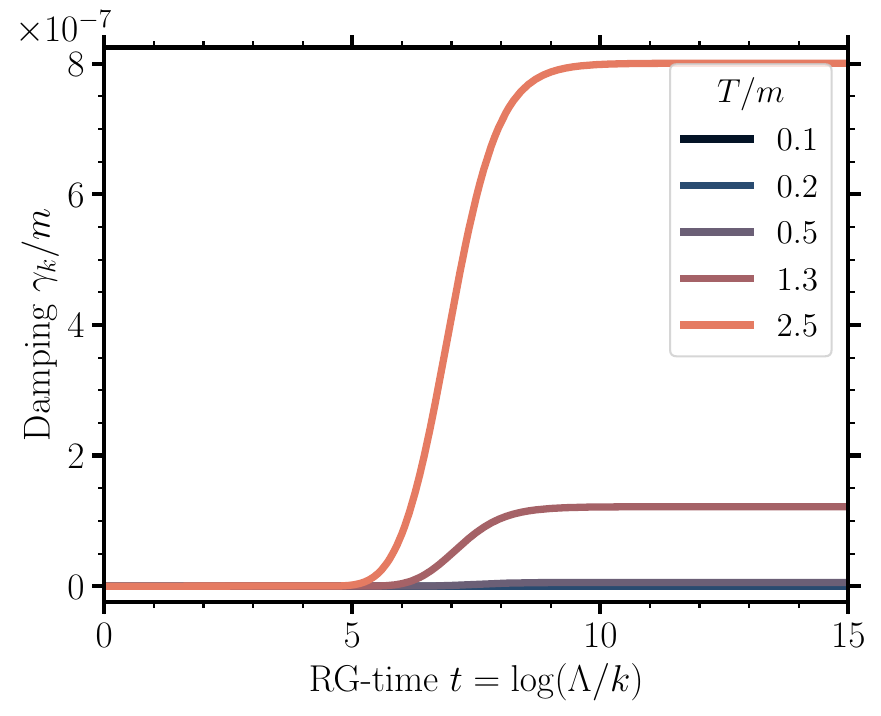}
    \caption{Solutions to the damping flow equation \eqref{eq:flow_gamma} at fixed coupling $\lambda=10^{-2}$ for different temperatures.}
    \label{fig:solution_damping_flow} 
\end{figure}
\begin{figure}[htbp]
    \centering 
    \includegraphics[width=0.9\columnwidth]{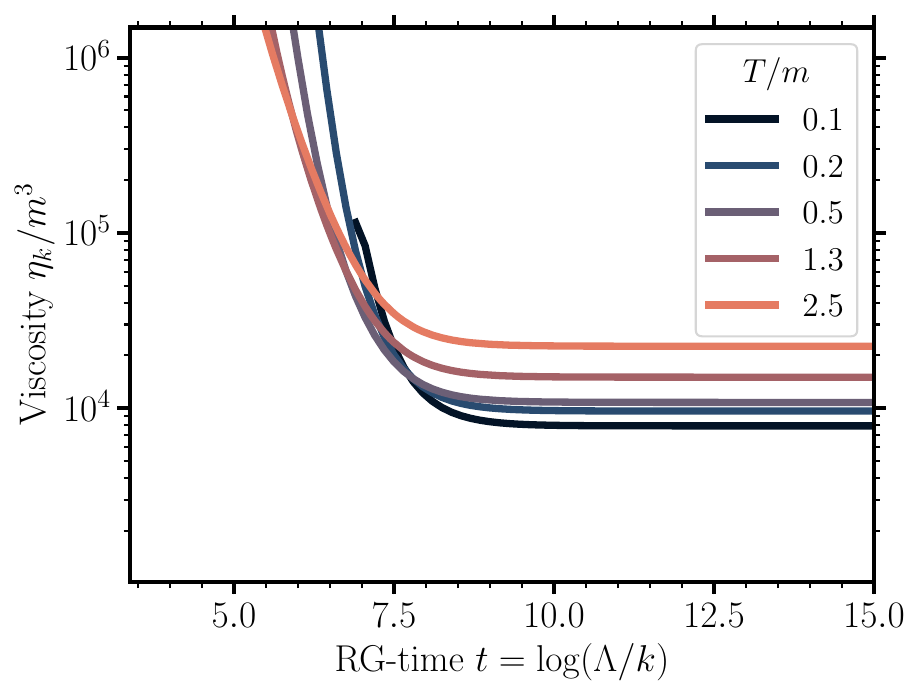}
    \includegraphics[width=0.9\columnwidth]{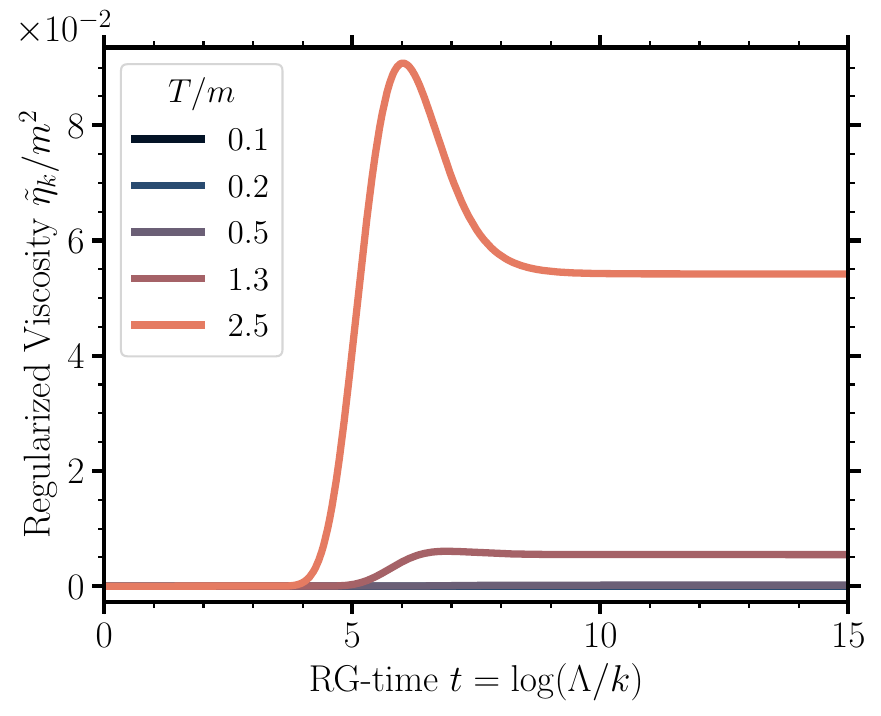}
    \caption{Solution of the viscosity coefficient (left) and the regularized viscosity (right) at $\lambda=10^{-2}$. 
    At early RG times the \textrm{Float64} precision limits the resolution of $\eta_k$ because $\gamma_k\approx 0 $ in that region.
    However, $\tilde\eta_k$ is well-behaved and close to zero.} 
    \label{fig:solution_eta_flow}
\end{figure}
Once the coupling $\kappa_k$ is non-vanishing, the flow of the Landau damping $\gamma_k$ is sourced. 
The latter is shown in \cref{fig:solution_damping_flow}.
At fixed interaction $\lambda$, we see that the damping increases monotonically with the temperature $T$.
Furthermore, the damping coefficient is far smaller than the mass in the weakly interacting regime, which is to be expected since the flow of $\gamma_k$ scales linearly with the coupling $\kappa_k$.
We can also see, similar to the flow of $\kappa_k$, that most dynamics arise when $k$ is close to the temperature.

A non-vanishing damping coefficient generates a non-vanishing flow of the regularized viscosity and a well-defined flow of the viscosity coefficient.
Both are shown in \cref{fig:solution_eta_flow}.  
As expected, the regularized viscosity is well-defined at the UV, being zero until thermal effects are being resolved by the scale $k$.  
The coefficient $\tilde{\eta}_k$ does not grow monotonically as it was the case in the previously mentioned coefficients. 
The viscosity flow equation \eqref{eq:flow_eta} suggests that the coefficient $\eta_k$ is expected to be a monotonically decreasing function with increasing RG time.
This can also be seen in the solution of $\eta_k$, given in \cref{fig:solution_eta_flow}, which shows the expected $\eu^{\beta k}$ behaviour at early RG-times.
Once the flow of $\gamma_k$ starts to fade-out, the flow of $\eta_k$ is still large enough to change the sign of $\partial_t \tilde \eta_k$, only to freeze out afterwards. 
The hydrodynamic transport coefficient $\eta$ can be reconstructed using the IR values of $\gamma_k$ and $\tilde \eta_k$ and its dependence on the temperature $T$ and interaction strength $\lambda$ will be shown in the next section. 

\label{ch:numeric_and_results}





\section{Results}\label{ch:results}
After discussing the flow equations and their solutions, we will discuss how the Landau damping and the shear viscosity depend on the temperature and interaction strength. 
\subsection{Landau Damping}\label{ch:results_landaudamping}
\begin{figure}[htbp]
    \centering 
    \includegraphics[width=0.9\columnwidth]{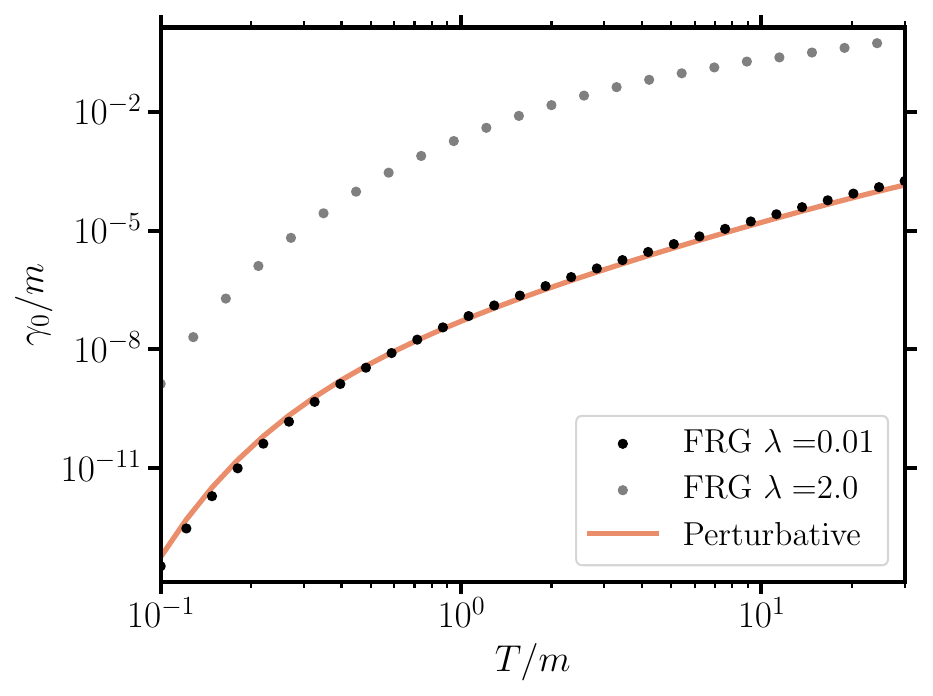}
    \caption{Landau damping $\gamma_0$ as a function of temperature at fixed interaction $\smash{\lambda=10^{-2}}$ and $\lambda=2$. 
    Each point is a solution to the flow equation \eqref{eq:flow_gamma}. 
    The solid line shows the perturbative result from~\cite{Jeon_1995}.}
    \label{fig:result_damping_temperature}
\end{figure}

The perturbative calculations from ref~\cite{Jeon_1995} suggest that the Landau damping is described, in the weakly interacting regime, by 
\begin{equation}
        \gamma^\text{pert}(T) = \frac{\lambda^2 T^2 }{2^8\pi^3 m_\mathrm{th}}\mathrm{Li}_2\left( \eu^{-\beta m_\mathrm{th}} \right),
        \label{eq:gamma_perturbative}
\end{equation}
where $\mathrm{Li}_n(z)$ is the polylogarithm of order $n$ and 
\begin{equation}
    m_\mathrm{th}^2 \approx m^2 + \frac{\lambda}{24}T^2 
    \label{eq:thermal_mass_perturbative}
\end{equation} 
is the perturbative thermal mass at one-loop\footnote{Note that ref~\cite{Jeon_1995} also included a cubic-interaction, which introduces a shift in the thermal mass. The expression here is consistent within our truncation with just a quartic coupling.} with the vacuum mass $m$. 
The FRG solution is shown in \cref{fig:result_damping_temperature}. 
We observe that the FRG solutions match quite accurately with the perturbative description if $\lambda \ll 1$ over several orders of magnitude in temperature.
In the high-relativistic limit, where $T\gg m$, the Landau damping scales linear with the temperature while the low-temperature regime shows the exponential scaling $\gamma\sim\eu^{-m/T}T^2/m$. 
Deviations from the perturbative results arises if the interaction strength is increased, as it is shown in \cref{fig:result_damping_temperature}.
While non-perturbative effects stay small in the non-relativistic regime, they become more transparent for larger temperatures where the contributions from the discontinuous vertex are most dominant.
 
\subsection{Viscosity}
\begin{figure}[htbp]
    \centering 
    \includegraphics[width=0.9\columnwidth]{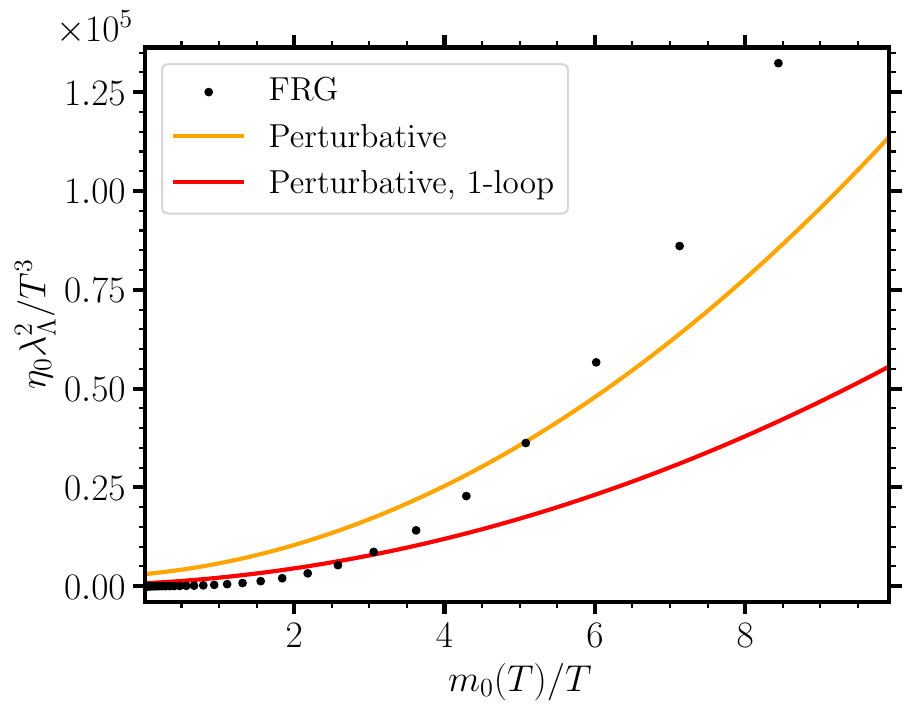}
    \caption{Shear viscosity coefficient $\eta_0$ over $T^3$ at $\Lambda=120 m$ and $\smash{\lambda=10^{-2}}$.
        The relativistic limit corresponds to $\smash{m_0(T)/T\rightarrow 0}$, and the non-relativistic case to $m_0(T)/T \gg 1$.
        Solid lines indicate fits from ref~\cite{Jeon_1995}.}
    \label{fig:result_eta_of_temperature}
\end{figure}

\begin{figure}
    \centering 
    \includegraphics[width=0.9\columnwidth]{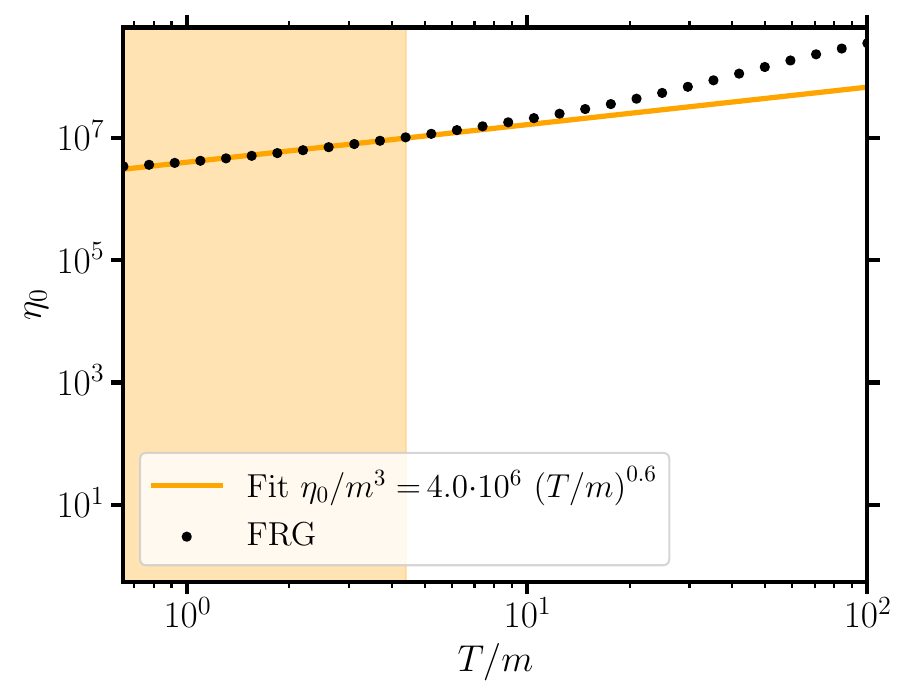}
    \caption{Low-temperature scaling behaviour of $\eta$ with respect to the temperature $T$. 
    The coloured region shows the interval where the fit has been taken.}
    \label{fig:result_eta_low_temperature}
\end{figure}

The viscosity coefficient $\eta_0(T, \lambda)$ is constructed from the regularized viscosity since $\eta_0(T) = \tilde{\eta}_0(T)/\gamma_0(T)$.
Similar to ref~\cite{Jeon_1995}, we can study the dependence of the dimensionless quantity $\eta\lambda^2 T^{-3}$ with respect to the ratio $m_0(T)/T$, which is shown in \cref{fig:result_eta_of_temperature}, where $m_0$ is the physical mass at the IR scale $k=0$.
At a fixed coupling $\lambda=10^{-2}$, we recover the expected scaling behaviour $\eta\sim T^3$ in the relativistic limit where $m_0(T)/T\ll 1$.
Compared to the perturbative results~\cite{Jeon_1995}, where it was found that 
\begin{equation}
    \begin{split}
        \eta_\text{pert}(T) \approx\, &3040\, T^3\lambda^{-2}\times\\
        &\left( 1 + 0.596 m_\mathrm{th}/T + 0.310m_\mathrm{th}^2/T^2 \right),
    \end{split}
    \label{eq:eta_pert_fit}
\end{equation} 
with $m_\mathrm{th}$ given in \cref{eq:thermal_mass_perturbative}, we find that the FRG solution deviates from the perturbative results. 
In the relativistic regime, the viscosity coefficient is smaller, while growing faster than the perturbative one in the non-relativistic regime.
The difference between the two results in the relativistic limit might occur, because we do not use the 'ladder-resummation' that was adapted in ref~\cite{Jeon_1995} and is closely connected to the wave function renormalization $Z_k(\phi)$, which is set to one in our proof-of-principle calculation.
We also resolve the low temperature limit where $T\ll m$ in \cref{fig:result_eta_low_temperature}. 
While \cref{eq:eta_pert_fit} does not recover the expected scaling $\smash{\eta\sim\sqrt{T}}$, which is the usual behaviour of non-relativistic gases encountered in the literature, see for example~\cite{schwabel_2006}, the non-perturbative calculation manages to resolve the non-relativistic limit more accurately.

\begin{figure}[htbp]
    \centering 
    \includegraphics[width=0.9\columnwidth]{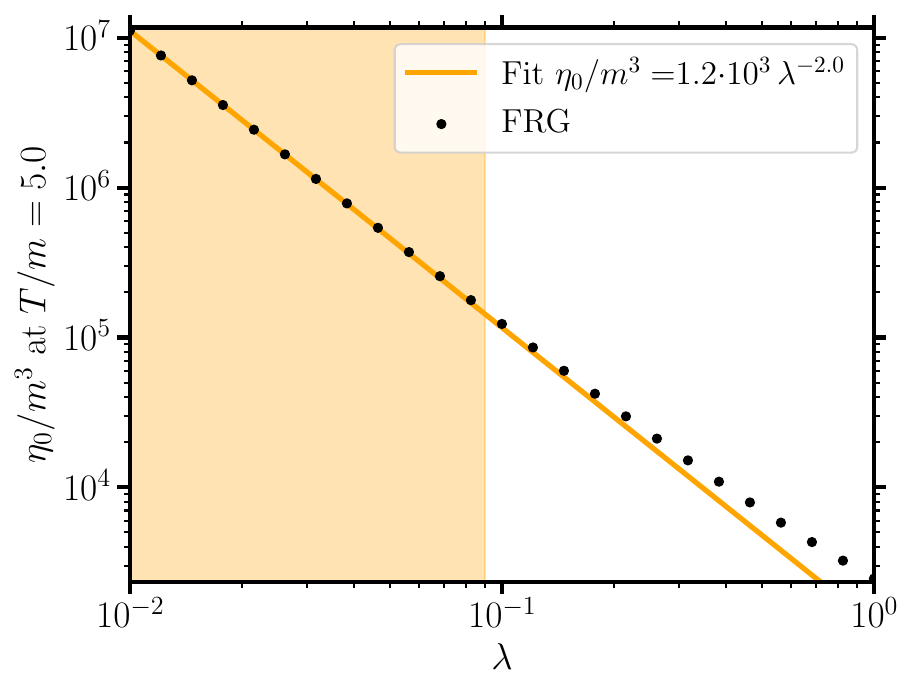}
    \caption{Dependence of the shear viscosity on the interaction strength $\lambda$ at $T=5 m$ and $\Lambda=120 m$.
            The shaded region indicates the interval over which the linear fit was taken.}
    \label{fig:result_eta_of_coupling}
\end{figure}

Furthermore, the dependence on the interaction strength $\lambda$ can also be studied.
This is shown in \cref{fig:result_eta_of_coupling}.
Since the mean-free-path of particles in the medium increases as the interaction strength decreases, larger regions of the fluid can equilibrate via momentum exchange which results in a large viscosity coefficient and vice versa, a small transport coefficient for large couplings corresponding to strongly interacting matter.
As $\lambda\rightarrow 0$ the transport coefficient seems to diverge as $\lambda^{-2}$, where the mean free path $l_\mathrm{MFP}\sim\lambda^{-1}$ tends to infinity.
The viscosity coefficient deviates from this power law when the coupling is closer to unity. 
This deviation from the perturbative scaling law is increasing with the temperature, where the non-perturbative effects become more relevant at higher temperatures.

\section{Conclusion} 
We have discussed a calculation of the shear viscosity coefficient of a relativistic, real scalar field model using the non-perturbative RG flow equation. 

We have developed a minimal truncation of the flowing action, which includes thermal corrections contributing to Landau damping in the two- and four-point functions.
Dissipative coefficients have been introduced by inserting explicit branch-cut structures, in a way that allows to do all Matsubara sums analytically.
Using this truncation, we derived flow equations for the Landau damping coefficient, thermal mass and the shear viscosity.
We consider this minimal scheme as suitable for a proof-of-concept, not yet aiming for qualitative precision. 

We calculated the temperature dependence of the Landau damping coefficient, located in the self-energy, and analyzed the dependence of the shear viscosity coefficient on the temperature and interaction strength. 
We recover results similar to the perturbative analysis performed in ref~\cite{Jeon_1995} for the Landau damping in the weakly interacting limit. 
Furthermore, we recover the expected scaling in the high-relativistic limit~\cite{Jeon_1995} where $\eta\sim T^3$, and non-relativistic limit, where $\eta\sim T^{1/2}$~\cite{schwabel_2006}.
Additionally, the dependence on the coupling is resolved.
While recovering the quadratic dependence, expected in the weakly-interacting theory~\cite{Jeon_1995}, we also find corrections arising form the non-perturbative FRG scheme which become relevant for strongly interacting matter.

The proof-of-principle study presented here can be extended in several directions. 
While the most important steps have been done, in regard to conceptual development of a truncation which captures dissipative effects, a systematic expansion towards a more complete truncation could be considered. 
These can include corrections from the flow of the quartic coupling, the vacuum flow of the mass but also a non-trivial wave-function renormalization, which is related to the perturbative 'ladder resummation' appearing in the shear viscosity calculation.
Within an improved truncation, one can proceed and study several other fluid dynamical quantities that are of interest, e.g. a bulk viscosity or second order coefficients like shear or bulk relaxation times. 
Additionally, thermodynamic properties, like the pressure~\cite{BLAIZOT2011165}, speed of sound, entropy density or susceptibilities could also be accessed.
More importantly, it would be most interesting to expand the ideas presented here by promoting the toy model to a more realistic effective theory, describing for example QCD matter that is encountered in heavy ion-collisions. 
Of course, one could also apply the conceptual ideas behind our truncation of the effective average action to study conserved currents that are related to coefficients like conductivities or diffusion coefficients for example.

We believe that an extension of the functional renormalization group method as we have developed it here can find many interesting applications.

\begin{acknowledgements} 
    We would like to thank Lars Heyen for many helpful discussions.  
    We also thank Eduardo Grossi for some insights into ODE solvers in the JULIA language and also thank Florain Atteneder for his valuable input on function interpolations.
\end{acknowledgements}

\appendix 

\section{Threshold functions}\label{ch:appendix_thresholdfunctions}
A reoccurring integral is the expression 
\begin{equation}
    \begin{split}
        \mathcal{I}_m^n&(|\mathbf{p}_1|,\ldots,|\mathbf{p}_m|; y) = \int_{-\infty}^{\infty}\!\frac{(1+2\mathrm{n}_\mathrm{B}(z))z^n }{\prod_{j=1}^m\prod_{l=1}^4(z+y-z_l^{(j)})}\dd z,
    \end{split}
\end{equation}
first encountered in \cref{eq:def_threshold_function}. 
The quantity $z_l^{(j)}$ is the $l$-th pole of the spectral density \eqref{eq:spectral_function_truncation} evaluated at the momentum $\mathbf{p}_j$ such that 
\begin{equation}
    \begin{split}
        z_1(\mathbf{p}_j) =& -\sqrt{\Omega_{p_j}^2 - \frac{\gamma_k^2}{2} + \frac{\iu\gamma_k}{2}\sqrt{4\Omega_{p_j}^2-\gamma_k^2}},\\
        z_2(\mathbf{p}_j) =& \quad\,\sqrt{\Omega_{p_j}^2 - \frac{\gamma_k^2}{2} - \frac{\iu\gamma_k}{2}\sqrt{4\Omega_{p_j}^2-\gamma_k^2}},        
    \end{split}
\end{equation}
if $\gamma_k<2\Omega_{p_j}$, 
\begin{equation}
    \begin{split}
        z_1(\mathbf{p}_j) =& -\sqrt{\Omega_p^2 - \frac{\gamma_k^2}{2} + \frac{\gamma_k}{2}\sqrt{\gamma_k^2-4\Omega_p^2}}, \\
        z_2(\mathbf{p}_j) =& -\sqrt{\Omega_p^2 - \frac{\gamma_k^2}{2} - \frac{\gamma_k}{2}\sqrt{\gamma_k^2-4\Omega_p^2}},
    \end{split}
\end{equation}
if $\gamma_k > 2\Omega_{p_j}$ and $z_{1,2}(\mathbf{p}_j) = -\iu\Omega_{p_j}$ if $\gamma_k = 2\Omega_{p_j}$.
Again, as it was mentioned in the main text, the remaining two poles, for the momentum $p_j$, are conjugate to $z_1$ and $z_2$.

We assume $m,n$ being positive natural numbers and satisfying the inequality $n<4m-1$, making sure that the integral $\mathcal{I}_m^n$ converges.
Additionally, we assume that all poles $\smash{z_l^{(j)}}$are non-degenerate.
The case of degenerate poles, where for example one momentum equals another, can be calculated later on by taking the limit $\mathbf{p}_j\rightarrow \mathbf{p}_i$ for the respective momenta.

Furthermore, we can decompose the fraction into the form 
\begin{equation}
    \begin{split}
        &\frac{1}{\prod_{j=1}^m\prod_{l=1}^4(z + y - z_l^{(j)})} = \\
        &\qquad \qquad \sum_{j=1}^m\sum_{l=1}^{4}\alpha_l^{(j)} \frac{z + z_l^{(j)} - y}{z^2-(z_l^{(j)}-y)^2},
    \end{split}
\end{equation}
with 
\begin{equation} 
    \alpha_l^{(j)} = \frac{1 }{\prod_{h=1}^m\prod_{\substack{i=1 \\ (h,i)\neq (j,l)} }^4(z^{(j)}_l - z^{(h)}_i)}.
\end{equation}
The remainder can now be split into contributions from the vacuum and thermal parts, where the latter contains the Bose-distribution $\mathrm{n}_\mathrm{B}(z)$, such that 
\begin{equation}
    \begin{split}
        &\mathcal{I}_m^n(|\mathbf{p}_1|,\ldots, |\mathbf{p}_m|; y) = \sum_{j=1}^m\sum_{l=1}^{4}\alpha_l^{(j)} \\
        &\times\begin{cases}
            2 v^{n+1}(z_l^{(j)}) + 4 t^{n+1}(z_l^{(j)}),\quad n\text{ even}\\
            2 z_l^{(j)} v^n(z_l^{(j)}) + 4 z_l^{(j)} t^n(z_l^{(j)}),\quad n\text{ odd}
        \end{cases}.
    \end{split}
\end{equation}
The vacuum and thermal parts are defined by the integrals  
\begin{align}
    v^d(u) &= \int_0^\infty \frac{z^d }{z^2 - u^2 } \dd z,\\
    t^d(u) &= \int_0^\infty \mathrm{n}_\mathrm{B}(z) \frac{z^d }{z^2 - u^2 } \dd z ,
\end{align}
with the index $d$ being an odd, positive integer and $u$ being a complex number.
The vacuum parts can be expressed by hypergeometric functions $\,_2F_1$, 
\begin{equation}
    \begin{split}
        v^d(u) =& \lim_{\tilde\Lambda\rightarrow 0}-\frac{\tilde\Lambda^{d+1}}{2(d+1)z^2}\bigg[ \,_2F_1\left(1,d+1,d+2; \frac{\tilde\Lambda }{u}\right) \\
        +&\,_2F_1\left(1,d+1,d+2; -\frac{\tilde\Lambda }{u}\right) \bigg],
    \end{split}
\end{equation}
where $\tilde\Lambda>0$ is a finite cutoff.
To arrive at this expression, we have used the integral representation of the hypergeometric function $\,_{2}F_1$ where
\begin{equation}
\begin{split}
        &\frac{\Gamma(b)\Gamma(c-b)}{\Gamma(c)} \,_{2}F_1(a, b, c; z)  \\
        &=\int_0^1 x^{b-1}(1-x)^{c-b-1}(1-zx)^{-a} \dd x.
\end{split}
\end{equation}
If $n<4m-1$, the limit $\tilde\Lambda\rightarrow \infty$ can be taken after the sum over all $\alpha_l^{(j)}$ coefficients is performed. 

The thermal parts are solved by
\begin{equation} 
    \begin{split}
        t^d(u) =& \sum_{j=0}^{\frac{d-1}{2}-1} C_{d-2j-1}u^{2j}\\
        +& \frac{u^{d-1}}{2}\left( \ln\left( \frac{\iu u \beta }{2\pi }\right) - \frac{\pi }{\iu u \beta } - \psi^{(0 )}\left( \frac{\iu u \beta }{2\pi} \right) \right),
    \end{split}
\end{equation}
with $\smash{C_{k} = \frac{\zeta(k)\Gamma(k)}{\beta^{k}}}$, $\mathrm{Im}\, u < 0$ and $d$ being odd.
If $\mathrm{Im}\, u >0$, one has to replace $u\leftrightarrow -u$. 
To arrive at this result, we used the integral representation of the \textit{digamma function}~\cite{WhittakerWatson:2009} 
\begin{equation}
    \psi^{(0)}(z) = \ln(z) - \frac{1}{2z} -2\int_0^\infty \frac{1}{\eu^{2\pi t} -1 }\frac{t }{t^2+z^2} \dd t,
\end{equation}
defined for $\mathrm{Re}\, z > 0$, the decomposition 
\begin{equation}
    \frac{t^{2n}}{t^2 + z^2 } = (-1)^n\frac{z^{2n} }{t^2 + z^2 } + \sum_{j=0}^{n - 1 }t^{2(n-j-1)}z^{2j}(-1)^j,
\end{equation} 
and the integral representation of the zeta function 
\begin{equation}
    \zeta(s) = \frac{1}{\Gamma(s)}\int_0^\infty \frac{t^{s-1}}{\eu^t - 1}\dd t,
\end{equation}
for $\mathrm{Re} \,s>0$.

\section{Contour Integrals}\label{ch:appendix_contour_integrals}
We will show the analytic results of the encountered Matsubara sums in this work by following the procedure outlined in \cref{ch:flowequations}.

\subsection{'Bubble-diagram'}
The $\phi^2$-$\phi^2$ 'bubble' diagram 
\begin{equation*} 
    \bubbleannotations{0.5},
\end{equation*}
encountered in the flow equation of the coupling $\kappa_k(\mathbf{p}^2)$ corresponds to the Matsubara sum 
\begin{equation}
    \begin{split}
        \mathcal{M}_{\lambda\lambda}(\iu \omega_m, \mathbf{p}; \mathbf{q}) =& \sum_{\omega_n} f(\iu\omega_n),
    \end{split}
\end{equation}
where $\mathbf{p}=\mathbf{p}_1 + \mathbf{p}_2$ and
\begin{equation}
    \begin{split}
        f(\iu\omega_n) =& \lambda^2 G_k(\iu\omega_n,\mathbf{q})G_k(\iu\omega_n + \iu \omega_m, \mathbf{q} + \mathbf{p}).
    \end{split}
\end{equation}
The function $f(z)$ introduces two branch cuts from the propagator along the real line and parallel to the real line at $z=-\iu\omega_m$.
Using this information we can identify the sum with the integral 
\begin{equation}
    \begin{split}
        \mathcal{M}_{\lambda\lambda}&(\iu \omega_m, \mathbf{p}; \mathbf{q}) = \lambda^2\int_{-\infty}^\infty (1+2\mathrm{n}_\mathrm{B}(z)) \\ 
        &\times\frac{1}{2}[\rho_k(z,\mathbf{q})G_k(z + \iu\omega_m, \mathbf{q} + \mathbf{p})\\ 
        &+ \rho_k(z,\mathbf{q} + \mathbf{p})G_k(z - \iu \omega_m, \mathbf{q})] \frac{\dd z }{2\pi },
    \end{split}
\end{equation}
where we used $-2\mathrm{Disc}[G_k](z,\mathbf{p}) = \rho_k(z,\mathbf{p})$.
Continuing the frequency to $\iu\omega_m \rightarrow \omega + \iu \epsilon$ and taking the imaginary part and evaluating the last expression at $p_t = p_u = 0$ results in the expression 
\begin{equation}
    \begin{split}
        &\mathrm{Im}\,\mathcal{M}_{\lambda\lambda}(\omega, \mathbf{p}; \mathbf{q})|_{p_\mathrm{t} = p_\mathrm{u} = 0} \\
        &=\frac{\lambda^2}{2^3\pi } \int_{-\infty}^\infty (1+2\mathrm{n}_\mathrm{B}(z)) [\rho_k(z,\mathbf{q})\rho_k(z + \omega, \mathbf{q} + \mathbf{p}) \\
        &\qquad\qquad\qquad\qquad\quad- \rho_k(z - \omega,\mathbf{q})\rho_k(z,\mathbf{q} + \mathbf{p})].
    \end{split}
\end{equation}
Furthermore, taking the derivative with respect to the frequency at the origin, the expression can be reduced to 
\begin{equation}
    \begin{split}
        &\frac{\partial }{\partial \omega }\mathrm{Im}\, \mathcal{M}_{\lambda\lambda}(\omega, \mathbf{p};\mathbf{q})|_{p_\mathrm{t} = p_\mathrm{u} = 0, \omega=0} = -\frac{\lambda^2}{2^3 \pi }\\ 
        &\times \beta\frac{\partial }{\partial \beta }\int_{-\infty}^\infty (1+2\mathrm{n}_\mathrm{B}(z))\rho_k(z,\mathbf{q})\rho(z,\mathbf{q}+\mathbf{p}) \frac{\dd z}{z},
    \end{split}
\end{equation}
which corresponds to the threshold function 
\begin{equation}
    \begin{split}
        \frac{\partial }{\partial \omega } \mathrm{Im}\, \mathcal{M}_{\lambda\lambda}&(\omega, \mathbf{p};\mathbf{q})|_{p_\mathrm{t} = p_\mathrm{u} = 0, \omega=0}\\
        &=-\frac{\lambda^2\gamma_k^2}{2^3 \pi }\beta\frac{\partial }{\partial \beta }\mathcal{I}_2^1(|\mathbf{q}|,|\mathbf{q}+\mathbf{p}|;0).
    \end{split}
\end{equation}

\subsection{Real part of the tadpole diagram}
The flow of the mass $m_k$ depends on the real part of the tadpole diagram. 
In contrast to the Landau damping, we know get both vertex contributions, from the quartic coupling $\lambda$ and the discontinuous vertex $\mathcal{A}(z,\mathbf{p})$.
The first Matsubara sum corresponds to the diagram 
\begin{equation*}
    \selfenergyquarticcoupling{0.5},
\end{equation*}
with 
\begin{equation}
    \begin{split}
        \mathcal{M}_\lambda(\mathbf{q}) = \sum_{\omega_n} \lambda G_k(\iu\omega_n, \mathbf{q}),
    \end{split}
\end{equation}
depending only on the loop momentum $\mathbf{q}$.
The summand does only exhibit a branch cut on the real line, produced by the propagator.
Hence, we can represent the sum by the integral  
\begin{equation}
    \begin{split}
        \mathcal{M}_{\lambda}(\mathbf{q}) = \frac{\lambda }{4\pi }\int_{-\infty}^\infty (1+2\mathrm{n}_\mathrm{B}(z))\rho_k(z,\mathbf{q}) \dd z.
    \end{split}
\end{equation}
Using the definition \eqref{eq:def_threshold_function} and taking the real part we see that 
\begin{equation}
    \mathrm{Re}\,\mathcal{M}_\lambda(\mathbf{q}) = \frac{\lambda \gamma_k }{4\pi }\mathcal{I}_1^1(|\mathbf{q}|; 0).
\end{equation}
For the contributions of the second diagram, involving the discontinuous vertex, we can start using \cref{eq:tadpole_matsubara_sum_final_with_disc,eq:tadpole_mastubara_sum_euclidean} and continue the external frequency to the real axis. 
Taking now the real part of the integral yields with the expression
\begin{equation}
    \begin{split}
        \mathrm{Re}\, &\mathcal{M}_{\mathcal{A}}(0, \mathbf{0}; \mathbf{q}) = \\
        &\kappa_k(\mathbf{q}^2)\int_{-\infty}^\infty (1+2\mathrm{n}_\mathrm{B}(z))z \mathrm{Re}\, G_k^\mathrm{R}(z, \mathbf{q})\frac{\dd z}{\pi} ,
    \end{split}
\end{equation}
and since 
\begin{equation}
    \mathrm{Re}\, G_k^\mathrm{R}(z,\mathbf{q}) = \frac{2(\Omega_q^2 - z ^2)}{(z^2 - \Omega_q^2)^2 + \gamma_k^2 z^2 },
\end{equation}
we can identify 
\begin{equation}
    \begin{split}
        \mathrm{Re}\, \mathcal{M}_{\mathcal{A}}(0,\mathbf{0};\mathbf{q}) =  \frac{2\kappa_k(\mathbf{q}^2)}{\pi}\left( \Omega_q^2 \mathcal{I}_1^1(|\mathbf{q}|; 0) - \mathcal{I}_1^3(|\mathbf{q}|; 0) \right).
    \end{split}
\end{equation}
\section{Asymptotics}\label{ch:appendix_asymptotics}
To simplify the computations at high $k$ and resolve the correct UV limit of the shear viscosity flow we need to expand all flow equations around $\beta k \gg 1$ and solve the flows of both $\gamma_k$ and $\kappa_k$ as they are crucial for the evolution of the viscosity coefficient $\eta_k$.
Starting with the $\kappa_k$-flow we can approximate it by neglecting the damping coefficient $\gamma_k$ and thus find the solution
\begin{equation}
    \kappa_k(|\mathbf{p}|) \approx -\frac{1}{16\pi|\mathbf{p}|} n_\mathrm{B}\left( -\beta\sqrt{m_k^2+k^2+\mathbf{p}^2/4} \right),
\end{equation}
which is valid for $\gamma_k\approx 0$ and just given by the perturbative bubble-diagram expanded around $\omega=0$, agreeing with the results in ref~\cite{Weldon_2001}.

Proceeding with the damping coefficient, we start by approximating the Matsubara sum at vanishing damping, which results in 
\begin{equation}
\begin{split}
        \mathrm{Im}\,\mathcal{M}_\mathcal{A}^{\mathrm{R}}(m_k, \mathbf{0};\mathbf{q}) \overset{\gamma\rightarrow 0}{\longrightarrow}& \frac{\pi }{\Omega_q}\bigg[2m_k\coth(\Omega_q \beta/2) \\
        +&(m_k - \Omega_q)\coth((m_k-\Omega_q)\beta/2)\\
        -&(m_k + \Omega_q)\coth((m_k + \Omega_q)\beta/2)\bigg].
\end{split}
\end{equation}
We can now use the expression for $\kappa_k$ given in \cref{eq:kappa_gamma_asymptotics} and approximate $|\mathbf{q}|^2/4\approx |\mathbf{q}|^2$ in the loop integral of \cref{eq:flow_gamma}. 
The resulting momentum integral can be solved analytically by expanding around $k\beta \gg 1$ which gives for $k\gg m_k$ the approximate result 
\begin{equation}
    \partial_t \gamma_k|_{k\approx\Lambda} = \frac{3 k^2 \sinh(\beta m_k )}{2^5\pi^3 m_k}\eu^{-2\beta_k}
\end{equation}
such that the flow is asymptotically given by 
\begin{equation}
    \gamma_k \sim \frac{3 k \sinh(\beta m_k )}{2^4\beta m_k \pi^3 }\eu^{-2\beta k }.
\end{equation}

Utilizing our results we can finally calculate the asymptotics of the shear viscosity flow. 
Starting with the flow equation \eqref{eq:flow_eta} and expanding the Matsubara sum in leading order of the damping, using 
\begin{equation}
    \begin{split}
        \psi^{(1)}\left( \frac{\beta(\gamma - 2\iu \Omega_q )}{4\pi } \right)&\rightarrow \psi^{(1)}\left( -\frac{\iu\beta \Omega_q }{2\pi }\right) \\
        &\quad+ \frac{\beta \gamma_k }{4\pi }\psi^{(2)}\left( -\frac{\iu\beta \Omega_q }{2\pi } \right), \\
        \psi^{(2)}\left( \frac{\beta(\gamma_k -2\iu \Omega_q )}{4\pi }  \right)&\rightarrow \psi^{(2)}\left( -\frac{\iu\beta \Omega_q }{2\pi } \right).
    \end{split}
\end{equation}
Furthermore making use of the identity $\mathrm{Re}\psi^{(1)}(\iu x) = -(2x^2)^{-1} - \pi^2/2\mathrm{csch}^2(\pi x)$ the flow can be written as 
\begin{equation}
    \partial_t \eta_k \overset{\gamma_k\rightarrow 0 }{\longrightarrow} \frac{\beta }{160\pi \gamma_k }\tilde{\partial}_t \int_0^\infty \dd q q^6 \Omega_q^{-2}\mathrm{csch}^2\left( \Omega_q\beta/2 \right),
\end{equation}
with $\Omega_q^2=m_k^2+k^2+q^2$.
While the integral can be solved analytically as a linear combination of Struve and Bessel functions $\mathbf{L}_\nu(z)$ and $K_\nu(z)$ we only need the leading order contribution which results in 
\begin{equation}
    \begin{split}
        \partial_t \eta_k \approx & \frac{\beta }{160\pi\gamma_k }2k^2\frac{\partial}{\partial M^2}\big[ 2\pi M^5 \big( \beta M \big[ K_3(\beta M)\mathbf{L}_{-4}(\beta M) \\
        &+ K_4(\beta M)\mathbf{L}_{-3}(\beta M) \big] - 1 \big) \big],
    \end{split}
\end{equation}
with $M^2=m_k^2+k^2$.
This equation can now be expanded around $M^2\approx k^2$ and $\beta k \gg 1$ to yield the asymptotic flow of the shear viscosity, used in \cref{eq:eta_tilde_asymptotics}
   
\newpage

\bibliography{inputs/sources.bib}

\end{document}